\documentclass[a4paper,11pt]{article}
\usepackage{amsmath, amssymb, graphics, graphicx}
\usepackage{tikz}
\usepackage[dvipsnames]{xcolor}
\usepackage{physics}
\usepackage{longtable}
\usepackage{subfigure}
\usepackage{multirow}
\usepackage{slashed}
\usepackage{mathtools}
\usepackage{musicography}
\usepackage{hyperref}
\usepackage{cite}

\usepackage{stackengine}
\stackMath

\newcommand{\mathsym}[1]{{}}

\newcommand{\rmd}{\mathrm{d}}
\newcommand{\rmi}{\mathrm{i}}
\newcommand{\rme}{\mathrm{e}}

\allowdisplaybreaks

\usepackage[shortlabels]{enumitem}

\begin{document}


\begin{titlepage}
\begin{center}
\vspace*{-1.0cm}

\vspace{1.2cm}

{\LARGE  {\fontfamily{lmodern}\selectfont \bf  From Galilei to Euclidean Carroll and the Alice Particle: The Times They Are a-Changin'}} \\[.2cm]

\vskip 1.3cm
{\large {\bf Eric A. Bergshoeff$^{\, \musHalf}$, Luca Romano$^{\, \musQuarter}$, Jan Rosseel$^{\, \musEighth}$, Elena Sim\'{o}n F\'{e}lix$^{\, \musQuarter}$ \\ and Sara Zeko$^{\, \musEighth \, \musSixteenth}$}}\\
\vskip 1cm

\begin{small}
{}$^{\musHalf}$ \textit{Van Swinderen Institute, University of Groningen\\ Nijenborgh 3, 9747 AG Groningen, The Netherlands}

\vspace{4mm}

{}$^{\musQuarter}$ \textit{Departamento de Electromagnetismo y Electr\'{o}nica, Universidad de Murcia \\ Campus de Espinardo, 30100 Murcia, Spain}

\vspace{4mm}

{}$^{\musEighth}$ \textit{Division of Theoretical Physics, Rudjer Bo\v{s}kovi\'c Institute \\
Bijeni\v{c}ka 54, 10000 Zagreb, Croatia} \\

\vspace{4mm}

{}$^{\musSixteenth}$ \textit{Department of Physics, Faculty of Science, University of Zagreb \\
10000 Zagreb, Croatia} \\

\vspace{3mm}

{\texttt{e.a.bergshoeff[at]rug.nl, lucaromano2607[at]gmail.com, jan.rosseel[at]irb.hr, elenapaola.simonf[at]um.es, sara.zeko[at]irb.hr}}

\end{small}

\end{center}

\vskip 0.7 cm
\begin{abstract}
\vskip.2cm\noindent

In generalized, also known as $p$-brane Galilei limits, the speed of light $c$ becomes infinite in the directions transverse to a $(p+1)$-dimensional Lorentzian worldvolume. In this paper, we explain that allowing the worldvolume to be Euclidean, and thus time to be transversal, $p$-brane Galilei limits turn into Carrollian ones, that we refer to as ``Euclidean $p$-brane Carroll limits", in which $c$ goes to zero in the $p+1$ worldvolume directions. This leads to a unified approach to taking Galilean and Carrollian limits, whose consequences we explore for $p=0$. We show that the spacetime symmetries that arise from the Euclidean 0-brane Carroll limit can be centrally extended to what we will call the Alice algebra, similar to how the Bargmann algebra centrally extends the Galilei symmetries. This gives rise to the novel notion of an Alice particle, and we obtain the Bargmann and Alice particle actions from a unified limit of the action of a relativistic massive particle or tachyon, suitably coupled to a one-form gauge potential. In the presence of a cosmological constant, we find that the Bargmann and Alice particles undergo stable motion for negative and positive cosmological constant, respectively. Finally, we show that the Bargmann and Alice particle actions can be obtained from null reduction of a massless particle action in a relativistic spacetime with one and two times. Our results indicate that in 10 dimensions, the Alice particle is a decoupling limit of a D$0{}^*$-brane in Hull's type IIA${}^*$ theory, similar to how the Bargmann particle is related to a D$0$-brane in type IIA string theory.

\end{abstract}

\end{titlepage}


%

\section{Introduction}

Although the status of timelike T-duality is not as settled as that of its spacelike counterpart, it has been known for some time that it serves as a gateway into an exotic corner of string and M-theory \cite{Hull:1998vg,Hull:1998ym,Hull:1998fh}. Passing from conventional string theories through the mirror that timelike T-duality provides, one enters into a looking-glass land of string and supergravity theories with bizarre characteristics, such as ghosts and unusual spacetime signatures. Despite (or perhaps because of) the subtleties that surround timelike reductions and dualities, the theories in this looking-glass land also have attractive features. For instance, the presence of ghosts in some of them evades well-known no-go theorems on the existence of de Sitter vacua in string theory \cite{Gibbons:1984kp,Maldacena:2000mw,Gibbons:2003gb}. Consequently, de Sitter vacua that preserve a twisted form of supersymmetry are as natural in looking-glass land as supersymmetric anti-de Sitter vacua are in the world of ordinary string theories. Moreover, timelike T-duality can give rise to supersymmetric branes, whose near-horizon geometry is a product of de Sitter space and a hyperboloid and whose worldvolume is Euclidean \cite{Hull:1998vg,Hull:1998ym,Hull:1998fh,Gutperle:2002ai}. Climbing through timelike T-duality's looking-glass thus allows one to explore holographic dualities between de Sitter gravity and Euclidean conformal field theory \cite{Hull:1998vg,Strominger:2001pn,Balasubramanian:2001nb}.

Recently, timelike T-duality has resurfaced in the context of non-Lorentzian decoupling limits of string and M-theory. One example of such a limit consists of taking the speed of light $c$ to infinity in the directions transverse to a string that winds around a compact direction. In this so-called nonrelativistic string limit, string excitations with negative winding numbers decouple, and the remaining ones with positive winding numbers obey nonrelativistic dispersion relations \cite{Gomis:2000bd,Danielsson:2000gi,Danielsson:2000mu}. The application of spacelike U-duality transformations to this limit then uncovers a web of generalized Galilean decoupling limits of string and M-theory \cite{Blair:2023noj,Gomis:2023eav,Blair:2024aqz,Blair:2025prd,Guijosa:2025mwh}. Here, we use the phrase ``generalized Galilean limits" to indicate that they can be interpreted as $c \rightarrow \infty$ limits in the $D-p-1$ Euclidean directions transverse to a Lorentzian $p$-brane (with $D$ the spacetime dimension and $0 \leq p \leq D-2$) \cite{Garcia:2002fa,Brugues:2004an,Gomis:2004pw,Bergshoeff:2020xhv,Bergshoeff:2023rkk}. Remarkably, these limits have been argued to be equivalent to Maldacena's AdS/CFT near-horizon limits \cite{Blair:2024aqz,Blair:2025prd,Guijosa:2025mwh}. As a consequence, AdS/CFT can also be viewed as open/closed string duality in string theories on backgrounds with generalized nonrelativistic (Newton-Cartan) asymptotics. By allowing for timelike U-dualities, one can obtain analogous statements in the theories that populate the looking-glass land. Curiously, timelike U-dualities transform generalized Galilean limits into Carrollian ones \cite{Blair:2023noj,Gomis:2023eav,Blair:2024aqz,Blair:2025prd,Blair:2025nno,Argandona:2025jhg}, in which $c$ goes to zero in $(D-p-1)$ Euclidean directions (for $0 \leq p \leq D-2$) \cite{Cardona:2016ytk,Roychowdhury:2019aoi,Kluson:2017fam,Kluson:2022jxh,Bergshoeff:2020xhv,Bergshoeff:2023rkk}. In looking-glass land, such generalized Carrollian limits are equivalent to Maldacena-like dS/CFT near-horizon limits, and dS/CFT can be recast as open/closed duality for strings in backgrounds with Carrollian asymptotics \cite{Blair:2025nno,Argandona:2025jhg}.\,\footnote{For recent work on Carroll limits of the AdS/CFT correspondence, see \cite{Fontanella:2025tbs}.} 

Usually, the generalized Galilean and Carrollian limits are viewed as each other's opposite. However, the fact that they are connected by timelike U-duality suggests that they are more closely related than often intuited. In this work, we will make explicit that this is indeed the case and that they can be treated in a unified manner. We will see that all generalized Galilean and Carrollian limits can be obtained by choosing a $(p+1)$-dimensional ``worldvolume" in spacetime, rescaling the directions longitudinal to it with a dimensionless parameter $\omega$, and taking $\omega \rightarrow \infty$. Whether such a limit is Galilean or Carrollian depends on the nature of the chosen worldvolume. For a Lorentzian worldvolume, one gets a generalized Galilean limit, that is referred to as the ``$p$-brane Galilei limit". On the other hand, if the worldvolume is Euclidean, the corresponding limit is generalized Carrollian that we will call the ``Euclidean $p$-brane Carroll limit".\,\footnote{In this paper, we will adopt the convention that the worldvolume of a $p$-brane is $(p+1)$-dimensional, regardless of whether it is Lorentzian or Euclidean.} With reference to the title of this paper, we can thus say that changing time from being a longitudinal to a transversal direction turns a Galilean limit into a Carrollian one. As mentioned above, in ordinary string/M-theory, the typical supersymmetric branes are Lorentzian, while in looking-glass land their worldvolumes are Euclidean. Unifying non-Lorentzian limits in the way just described, then explains why Galilean and Carrollian limits are the natural ones to consider in ordinary string theory and in looking-glass land, respectively.

We will mainly illustrate the unified limit procedure for $p=0$, when the chosen worldvolume is a one-dimensional worldline. When this worldline lies along a timelike direction, the corresponding 0-brane Galilei limit is the standard Galilei limit, in which $c \rightarrow \infty$ in all spatial directions. The case where the worldline lies along a spacelike direction, and is thus tachyonic, defines the Euclidean $0$-brane Carroll limit, in which $c \rightarrow 0$ only along the chosen worldline direction. Note in particular that this is not the usual Carroll limit, for which instead $c \rightarrow 0$ in all spatial directions. However, as we will argue, there is a sense in which it is the Euclidean $0$-brane, and not the standard Carroll limit, that can be viewed as the natural Carrollian counterpart of the Galilei limit. In particular, our unified approach implies that the standard Galilei and Euclidean $0$-brane Carroll limits have similar properties, of which we will highlight two. First, we will see that both limits lead to symmetry algebras that can be centrally extended. Secondly, we will argue that systems that are invariant under both types of centrally extended symmetries can be obtained from a null reduction.

It is well-known that in nonrelativistic physics, Noether's theorem associates conservation of mass to an extra symmetry generator that centrally extends the Galilei algebra to the Bargmann algebra. Moreover, a worldline action for a massive ``Bargmann particle", that realizes the central charge in a non-trivial manner, can be obtained by taking a critical nonrelativistic $c \rightarrow \infty$ limit of the action of a massive relativistic particle, coupled to a one-form gauge potential. This limit is called critical because a divergent rest energy contribution is cancelled by a fine-tuned choice for the gauge potential. Physically, this cancellation has the effect of decoupling the anti-particles of the relativistic theory, leaving one with particles only. The usual Carroll algebra does not allow for a central extension and consequently, there is no corresponding Carroll particle action that is obtained from a critical limit. By contrast, we will see that our unified approach to taking non-Lorentzian limits shows that the Carrollian symmetries that arise from the Euclidean $0$-brane Carroll limit can be centrally extended to what we will call the ``Alice algebra". Taking a critical Euclidean $0$-brane Carroll limit of the worldline action of a relativistic tachyon, coupled to a gauge potential, then gives the action of an ``Alice particle", a novel Carrollian analogue of the Bargmann particle. We will show that the inclusion of a cosmological constant leads to a quadratic potential in the Bargmann \cite{Gibbons:2003rv,Brugues:2006yd} and Alice particle actions and that the Bargmann and Alice particles are stable for negative and positive cosmological constant, respectively.

Taking a critical 0-brane Galilei limit is not the only way in which one can obtain the Bargmann particle action and the nonrelativistic Newton-Cartan geometry to which it couples. These can also be obtained from a null reduction of the action of a massless particle moving in a Lorentzian background of one dimension higher \cite{Duval:1984cj}.  Such a reduction on a null circle is best understood as a limit of reductions on spacelike circles \cite{Seiberg:1997ad,Sen:1997we}. From this point of view, the central charge transformation arises from translations in the extra null direction. We will show that the Alice particle action can be obtained from a null reduction in a similar manner. In this case, however, the higher-dimensional starting point is the action of a massless particle that moves in a spacetime with two time directions and the null reduction is defined as a limit of reductions in which one of these times is compactified.\,\footnote{This procedure is different from recently proposed ways to obtain Carroll structures from null reduction \cite{Chen:2023pqf,Sharma:2025rug,Majumdar:2025juk,Majumdar:2026iol}. These proposals deal with metric structures that are locally invariant under the symmetries that arise from the usual Carroll limit. As pointed out above, these are different from the symmetries that the Euclidean $0$-brane Carroll limit leads to, whose null reduction origin we discuss in this paper.} Here, we then encounter a second way in which the title of our paper can be interpreted: one can pass from Bargmann to Alice by changing the number of times from one to two before doing a null reduction.

The Bargmann and Alice particles, that we discuss in this paper, play an important role in ordinary string theory and in the looking-glass land that arises from timelike U-duality. In a ten-dimensional target spacetime, the Bargmann particle action corresponds to a low-energy decoupling limit of the worldline action of a D0-brane in type IIA string theory. Its maximally supersymmetric generalization to multiple D0-branes then gives the BFSS matrix model that has been conjecturally proposed as a non-perturbative formulation of M-theory (i.e., strongly coupled type IIA string theory) \cite{Banks:1996vh,Susskind:1997cw}. The null reduction that gives the Bargmann particle, and in particular its interpretation as a limit of spacelike circle compactifications, lies at the heart of this conjecture. Indeed, in this limit, only the D0-branes that carry positive Kaluza-Klein momentum, remain as light degrees of freedom, whereas anti-D0-branes and all other type IIA states with negative or zero Kaluza-Klein momentum decouple. Looking-glass land contains a type IIA${}^*$ theory, that lives in signature $(1,9)$ and that is the timelike reduction of an M${}^*$ theory in spacetimes with signature $(2,9)$. This makes it tempting to surmise that the null reduction that is a limit of timelike circle compactifications and gives the Alice particle, acts as a decoupling limit that similarly only keeps D0${}^*$-branes with positive Kaluza-Klein momentum. A suitable generalization of the Alice particle action to include maximal supersymmetry and multiple D0${}^*$-branes is then expected to give a non-perturbative matrix model description of strongly coupled type IIA${}^*$ string theory \cite{Blair:2025nno,Argandona:2025jhg}.

The outline of this paper is as follows. In Section \ref{sec:unification}, we explain how all generalized $p$-brane Galilean and Carrollian limits can be unified by viewing the latter as Euclidean brane Carroll limits. In Section \ref{sec:lookingglass}, we physically interpret the four limits that have one longitudinal or transversal direction, and we discuss the Bargmann and Alice algebras that centrally extend the symmetries that arise from the $0$-brane Galilei and Euclidean $0$-brane Carroll limits. Section \ref{sec:BargmannAlice} is devoted to the derivation of the Bargmann and Alice particle actions from our unified limit procedure, both without and in the presence of a cosmological constant. In Section \ref{sec:nullred}, we provide a different derivation of these particles, namely from null reductions, that can be viewed as limits of reductions over spacelike and timelike circles, of a massless particle. We end with conclusions and an outlook on future work in Section \ref{sec:outlook}. Finally, this paper also contains two appendices that offer different perspectives on the Bargmann and Alice particles. In Appendix \ref{sec:SchrCarrSchr}, we show how the Schr\"odinger and Euclidean Carroll-Schr\"odinger equations that arise from quantization of these particles can be obtained from our unified limit. A similar discussion for the dispersion relations that are obeyed by the Bargmann and Alice particles is contained in Appendix \ref{sec:disprels}.

\section{A unified approach to Galilei and Carroll limits} \label{sec:unification}

In this section, we will describe how all generalized Galilean and Carrollian limits can be discussed in a way that is more unified than typically done. As an example, we will show what the usual Galilei and Carroll limits (in which $c$ goes to infinity and zero in all spatial directions) look like from this unified perspective. We will comment on how our treatment of Carrollian limits differs from most of the existing literature and motivate why we prefer to take a new point of view. 

We will be concerned with generalized Galilean and Carrollian limits of relativistic theories in spacetimes that can be flat or curved. In case we consider flat $D$-dimensional Minkowski spacetime, we will denote the coordinates of an inertial frame by $X^{A}$, $A=0,1,\dots, D-1$. They transform under rigid, infinitesimal Lorentz transformations with parameters $\Lambda^{A}{}_{B}$ as follows:
\begin{align}
    \delta X^{A} = \Lambda^{A}{}_{B} \, X^{B} \,.
\end{align}
When discussing limits of curved spacetimes, we will find it convenient to work with a Vielbein $E_{\mu}{}^{A}$ (with the flat index $A$ again taking the values $0,1,\cdots, D-1$) in terms of which the metric is given by 
\begin{align}
    \label{eq:metricvielb}
    G_{\mu\nu} = E_{\mu}{}^{A} E_{\nu}{}^{B} \eta_{A B} \,.
\end{align}
Here, we adopt the mostly plus signature convention for the Minkowski metric $\eta_{A B}$. The Vielbein transforms under infinitesimal local Lorentz transformations, whose parameters we will (with harmless abuse of notation) also denote by $\Lambda^{A}{}_{B}$, as follows:
\begin{align}
    \delta E_{\mu}{}^{A} = \Lambda^{A}{}_{B} \, E_{\mu}{}^{B} \,.
\end{align}
Before we describe the generalized Galilei and Carroll limits, let us briefly comment on dimensions. We take all coordinates $X^{A}$, including the timelike $X^0$ coordinate, to have dimensions of length. All Vielbein components, including the timelike one $E_{\mu}{}^0$, are assumed to be dimensionless. As a consequence, all Lorentz transformation parameters $\Lambda^{A}{}_{B}$ are also dimensionless.

The generalized Galilean and Carrollian limits can be treated in a unified manner as follows.  First, we partition the flat index $A$ into a ``longitudinal" index $A^\prime$, that takes on $p+1$ values, and a ``transversal" one $\tilde{A}$, that takes on the remaining $D-p-1$ values: 
\begin{align}
    A = (A^\prime, \tilde{A}) \,, \qquad \text{with} \ \ \ \ \text{range of } A^\prime = p+1 \ \ \ \ \text{and} \ \ \ \ \text{range of }
 \tilde{A} = D - p - 1 \,.
 \end{align}

\noindent As we shall discuss more at the end of this section, the terms "longitudinal" and "transversal" refer to their interpretation as indicating directions along and orthogonal to the worldvolume of a $p$-brane with respect to which we take the limit. Next, we introduce a dimensionless parameter $\omega$ and use it for the following redefinitions of the coordinates $X^{A}$, Vielbeine $E_{\mu}{}^{A}$ and Lorentz transformation parameters $\Lambda^{A}{}_{B}$:
\begin{align}
    \label{eq:uniflim}
      X^{A^\prime} &= \omega \, x^{A^\prime} \,,  & X^{\tilde{A}} &= x^{\tilde{A}} \,, \nonumber \\  E_{\mu}{}^{A^\prime} &= \omega\, e_{\mu}{}^{A^\prime} \,, & E_{\mu}{}^{\tilde{A}} &= e_{\mu}{}^{\tilde{A}} \,, \nonumber \\
     \Lambda^{A^\prime}{}_{B^\prime} &= \lambda^{A^\prime}{}_{B^\prime} \,, &  \Lambda^{\tilde{A}}{}_{\tilde{B}} &= \lambda^{\tilde{A}}{}_{\tilde{B}} \,, &   \Lambda^{A^\prime}{}_{\tilde{A}} &= \frac{1}{\omega} \lambda^{A^\prime}{}_{\tilde{A}} \,.
\end{align}
Here, it is understood that in case we consider limits in a fixed (e.g., flat) background spacetime, we rescale only the coordinates and Lorentz transformation parameters, whereas in arbitrary curved spacetimes, the above redefinitions are only applied to the Vielbeine and Lorentz transformation parameters. Finally, after applying the redefinitions \eqref{eq:uniflim}, we take the $\omega \rightarrow \infty$ limit. 

Note that up to this point, we have been careful not to specify whether the time direction is longitudinal or transversal. It is this choice that determines whether the limit is of the Galilean or Carrollian type. In particular, when the time direction is longitudinal, the above procedure yields a generalized Galilean limit, that is called the $p$-brane Galilei limit. When time is a transversal direction and all longitudinal directions are thus Euclidean, one obtains a generalized Carrollian limit, that we will refer to as the Euclidean $p$-brane Carroll limit. To justify this, let us look at the limit of the transformations of the coordinates $X^{A}$ and Vielbein fields $E_{\mu}{}^{A}$ under generalized boosts, i.e., the Lorentz transformations with parameters $\Lambda_{A^\prime}{}^{\tilde{A}} \ \ (=-\Lambda^{\tilde{A}}{}_{A^\prime})$  that mix longitudinal and transversal directions:
\begin{align}
    \label{eq:relboosts}
    \delta X^{A^\prime} = \Lambda^{A^\prime}{}_{\tilde{A}} \, X^{\tilde{A}} \,, \quad \delta X^{\tilde{A}} = -\Lambda_{A^\prime}{}^{\tilde{A}} \, X^{A^\prime} \,, \qquad \quad  \delta E_{\mu}{}^{A^\prime} = \Lambda^{A^\prime}{}_{\tilde{A}} \, E_{\mu}{}^{\tilde{A}} \,, \quad \delta E_{\mu}{}^{\tilde{A}} = -\Lambda_{A^\prime}{}^{\tilde{A}} \, E_{\mu}{}^{A^\prime} \,.
\end{align}
After applying the redefinitions \eqref{eq:uniflim}, the $\omega \rightarrow \infty$ limit of these transformation rules is given by:
\begin{align}
    \label{eq:unifboost}
   \delta x^{A^\prime} = 0 \,, \qquad \delta x^{\tilde{A}} = -\lambda_{A^\prime}{}^{\tilde{A}} \, x^{A^\prime} \,, \qquad \qquad \qquad \delta e_{\mu}{}^{A^\prime} = 0 \,, \qquad \delta e_{\mu}{}^{\tilde{A}} = - \lambda_{A^\prime}{}^{\tilde{A}} \, e_{\mu}{}^{A^\prime} \,.
\end{align}
When the time direction is longitudinal, the first boost rule of \eqref{eq:unifboost} sends the spatial coordinates $x^{\tilde{A}}$ to the $x^{A^\prime}$ that contain the timelike coordinate. This boost transformation behaviour, that can be schematically summarized as follows:
\begin{align}
    \text{spatial } x^{\tilde{A}} \ \ \ \rightarrow \ \ \ x^{A^\prime} \ni \text{time} \ \ \ \rightarrow \ \ \ 0 \,,
\end{align}
indicates that the coordinates $x^{\tilde{A}}$, $x^{A^\prime}$ form an indecomposable, reducible representation of a Galilean symmetry algebra. Likewise, since $e_{\mu}{}^{A^\prime}$ has a component along the time direction, it can be interpreted as a generalized clock. The second rule of \eqref{eq:unifboost} then boosts the generalized ruler $e_{\mu}{}^{\tilde{A}}$ to this clock, as is appropriate in a geometry with a generalized Galilean structure group. In case the time direction is transversal, the boosts \eqref{eq:unifboost} of the coordinates $x^{A^\prime}$ and $x^{\tilde{A}}$ are Carrollian, since the timelike coordinate is transformed to spatial ones. These coordinates form an indecomposable, reducible representation of a (generalized) Carrollian symmetry algebra, whose action is schematically given by:
\begin{align}
    x^{\tilde{A}} \ni \text{time} \ \ \ \rightarrow \ \ \ \text{spatial } x^{A^\prime} \ \ \ \rightarrow \ \ \ 0 \,.
\end{align}
Moreover, $e_{\mu}{}^{\tilde{A}}$ is now interpreted as a generalized clock that is boosted to the generalized ruler $e_{\mu}{}^{A^\prime}$, as is the case in Carrollian geometry. In summary, by performing the single set of rescalings \eqref{eq:uniflim}, but allowing time to be rescaled differently, all generalized Galilean and Carrollian limits can be unified as follows:
\begin{align}
    \label{eq:uniflim2}
    \text{$p$-brane Galilei limit: } \ \ \ & \text{redefinitions \eqref{eq:uniflim} with time } \in A^\prime \ \ \ \text{and} \ \ \ \omega \rightarrow \infty \,, \nonumber \\
     \text{Euclidean $p$-brane Carroll limit: } \ \ \ & \text{redefinitions \eqref{eq:uniflim} with time }  \in \tilde{A} \ \ \ \text{and} \ \ \ \omega \rightarrow \infty \,.
\end{align}

Let us, by way of example, show how the usual Galilei and Carroll limits fit in the above-described scheme. The former is what we just called the $0$-brane Galilei limit. There is thus one longitudinal direction, that lies along the timelike coordinate $X^0$ ($A^\prime = 0$), whereas the remaining spatial coordinates $X^{a} = X^{\tilde{A}}$ ($a = \tilde{A} = 1, \cdots, D-1$) are transversal. The redefinitions \eqref{eq:uniflim} of these coordinates and Lorentz transformation parameters are as follows:
\begin{alignat}{3}
\label{eq:Gallimit}
 & \text{Galilei limit:} \qquad \ & X^0 &= \omega\, x^0 \equiv \omega \, c \, t \,, \qquad \qquad \qquad & X^{a} &= x^{a} \,, \nonumber \\ & \text{(= $0$-brane Galilei limit)} \qquad & \Lambda^{0}{}_{a} &= \frac{1}{\omega} \lambda^{0}{}_{a} \equiv \frac{1}{\omega \, c} \lambda_{a} \,, \qquad & \Lambda^{a}{}_{b} &= \lambda^{a}{}_{b} \,.
\end{alignat}
Here, we have redefined the Galilean timelike coordinate $x^0$ and boost parameters $\lambda^{0}{}_{a}$ with a power of $c$,
so that the transformation rule \eqref{eq:unifboost} takes the standard form of a Galilean boost:
\begin{align}
    \delta x^{a} = \lambda^{a} \, t \,, \qquad \qquad \qquad \delta t = 0 \,.
\end{align}
The parameters $\lambda^{a}$ are no longer dimensionless, but have the dimension of length/time (velocity). 

The usual Carroll limit, in which $c$ goes to zero in all spatial directions, corresponds to the Euclidean $(D-2)$-brane Carroll limit. In this case, the time coordinate is the single transversal coordinate $X^0$ ($\tilde{A} = 0$) and the longitudinal coordinates $X^{a} = X^{A^\prime}$ ($a = A^\prime = 1,\cdots, D-1$) are all spacelike. The redefinitions \eqref{eq:uniflim} of the coordinates and the Lorentz transformation parameters now read: 
\begin{alignat}{3}
\label{eq:Carrlimit}
    & \text{Carroll limit:}  \qquad \qquad & X^{a} &= \omega \, x^a \,, \qquad \qquad & X^{0} &= x^0 \equiv c\, t \,, \nonumber \\ & \text{(= Eucl. $(D-2)$-brane Carroll limit)} \qquad & \Lambda^{0}{}_{a} &= \frac{1}{\omega} \lambda^{0}{}_{a} \equiv \frac{c}{\omega} \tilde{\lambda}^{a} \,, \qquad & \Lambda^{a}{}_{b} &= \lambda^{a}{}_{b} \,.
\end{alignat}
Once again, we have redefined $x^0$ and the boost parameters $\lambda^{0}{}_{a}$ with a power of $c$,
so that \eqref{eq:unifboost} becomes the standard Carroll boost rule:
\begin{align}
    \delta t = \tilde{\lambda}_{a} \, x^{a} \,, \qquad \qquad \qquad \delta x^{a} = 0 \,,
\end{align}
where the boost parameters $\tilde{\lambda}^{a}$ now have the dimension of time/length (inverse velocity). Referring to which spacetime directions grow with $\omega$, we can summarize the difference between the Galilei \eqref{eq:Gallimit} and the Carroll limit \eqref{eq:Carrlimit}, by saying that in the former ``time beats space", whereas in the latter ``space beats time".

\begin{figure}
    \centering
\resizebox{\textwidth}{!}{

\tikzset{every picture/.style={line width=0.75pt}} 

\begin{tikzpicture}[x=0.75pt,y=0.75pt,yscale=-1,xscale=1]

\draw  [color={rgb, 255:red, 0; green, 0; blue, 0 }  ,draw opacity=1 ][fill={rgb, 255:red, 205; green, 205; blue, 205 }  ,fill opacity=1 ] (50,100) -- (500,100) -- (500,240) -- (50,240) -- cycle ;
\draw  [fill={rgb, 255:red, 255; green, 255; blue, 255 }  ,fill opacity=1 ] (60,110) -- (210,110) -- (210,230) -- (60,230) -- cycle ;
\draw  [fill={rgb, 255:red, 182; green, 183; blue, 255 }  ,fill opacity=1 ] (350,110) -- (490,110) -- (490,230) -- (350,230) -- cycle ;
\draw  [fill={rgb, 255:red, 208; green, 2; blue, 27 }  ,fill opacity=1 ][line width=0.75]  (325,155) -- (254.11,155) -- (254.11,140) -- (220,170) -- (254.11,200) -- (254.11,185) -- (325,185) -- cycle ;\draw  [fill={rgb, 255:red, 208; green, 2; blue, 27 }  ,fill opacity=1 ][line width=0.75]  (340,155) -- (337,155) -- (337,185) -- (340,185) -- cycle ;\draw  [fill={rgb, 255:red, 208; green, 2; blue, 27 }  ,fill opacity=1 ][line width=0.75]  (334,155) -- (328,155) -- (328,185) -- (334,185) -- cycle ;

\draw  [color={rgb, 255:red, 0; green, 0; blue, 0 }  ,draw opacity=1 ][fill={rgb, 255:red, 205; green, 205; blue, 205 }  ,fill opacity=1 ] (882.5,100) -- (1332.5,100) -- (1332.5,240) -- (882.5,240) -- cycle ;
\draw  [fill={rgb, 255:red, 255; green, 255; blue, 255 }  ,fill opacity=1 ] (892.5,110) -- (1042.5,110) -- (1042.5,230) -- (892.5,230) -- cycle ;
\draw  [fill={rgb, 255:red, 182; green, 183; blue, 255 }  ,fill opacity=1 ] (1182.5,110) -- (1322.5,110) -- (1322.5,230) -- (1182.5,230) -- cycle ;
\draw  [fill={rgb, 255:red, 208; green, 2; blue, 27 }  ,fill opacity=1 ][line width=0.75]  (1067.5,155) -- (1138.39,155) -- (1138.39,140) -- (1172.5,170) -- (1138.39,200) -- (1138.39,185) -- (1067.5,185) -- cycle ;\draw  [fill={rgb, 255:red, 208; green, 2; blue, 27 }  ,fill opacity=1 ][line width=0.75]  (1052.5,155) -- (1055.5,155) -- (1055.5,185) -- (1052.5,185) -- cycle ;\draw  [fill={rgb, 255:red, 208; green, 2; blue, 27 }  ,fill opacity=1 ][line width=0.75]  (1058.5,155) -- (1064.5,155) -- (1064.5,185) -- (1058.5,185) -- cycle ;

\draw  [fill={rgb, 255:red, 195; green, 173; blue, 173 }  ,fill opacity=1 ] (520.25,170) -- (556.99,135) -- (556.99,154.6) -- (823.51,154.6) -- (823.51,135) -- (860.25,170) -- (823.51,205) -- (823.51,185.4) -- (556.99,185.4) -- (556.99,205) -- cycle ;

\draw (420,170) node  [font=\Large] [align=left] {\begin{minipage}[lt]{80.3pt}\setlength\topsep{0pt}
\begin{center}
\textbf{Transverse}\\$D-p-1$\\Euclidean
\end{center}

\end{minipage}};
\draw (135,170) node  [font=\Large] [align=left] {\begin{minipage}[lt]{90.83pt}\setlength\topsep{0pt}
\begin{center}
\textbf{Longitudinal}\\$p+1$\\Lorentzian
\end{center}

\end{minipage}};
\draw (278.5,170) node  [font=\Large,color={rgb, 255:red, 255; green, 255; blue, 255 }  ,opacity=1 ] [align=left] {\begin{minipage}[lt]{40.28pt}\setlength\topsep{0pt}
\begin{center}
Boost
\end{center}

\end{minipage}};
\draw (1252.5,170) node  [font=\Large] [align=left] {\begin{minipage}[lt]{90.83pt}\setlength\topsep{0pt}
\begin{center}
\textbf{Longitudinal}\\$p+1$\\Euclidean
\end{center}

\end{minipage}};
\draw (967.5,170) node  [font=\Large] [align=left] {\begin{minipage}[lt]{80.3pt}\setlength\topsep{0pt}
\begin{center}
\textbf{Transverse}\\$D-p-1$\\Lorentzian
\end{center}

\end{minipage}};
\draw (1111,170) node  [font=\Large,color={rgb, 255:red, 255; green, 255; blue, 255 }  ,opacity=1 ] [align=left] {\begin{minipage}[lt]{40.28pt}\setlength\topsep{0pt}
\begin{center}
Boost
\end{center}

\end{minipage}};
\draw (690.25,170) node  [font=\Large] [align=left] {\begin{minipage}[lt]{47.62pt}\setlength\topsep{0pt}
\begin{center}
Duality
\end{center}

\end{minipage}};
\draw (1107.5,72.5) node  [font=\huge] [align=left] {\begin{minipage}[lt]{308.93pt}\setlength\topsep{0pt}
\begin{center}
\textbf{ \ \ \ Euclidean $\boldsymbol{p}$-brane Carroll \ \ \ }
\end{center}

\end{minipage}};
\draw (275,72.5) node  [font=\huge] [align=left] {\begin{minipage}[lt]{175.19pt}\setlength\topsep{0pt}
\begin{center}
\textbf{ \ $\boldsymbol{p}$-brane Galilei \ }
\end{center}

\end{minipage}};

\end{tikzpicture}
}
    \caption{Duality between the $p$-brane Galilei and Euclidean $p$-brane Carroll limits.}
    \label{fig:diag1}
\end{figure}
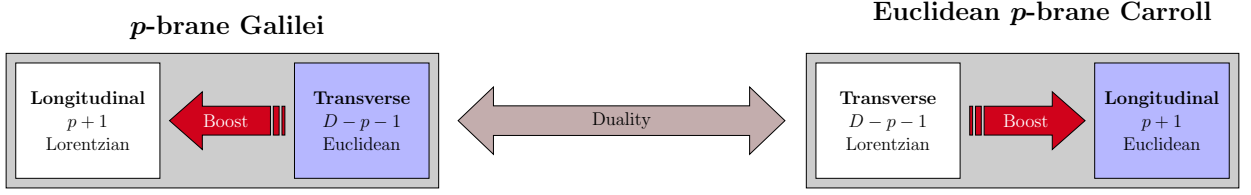

Our unified treatment of $p$-brane non-Lorentzian limits agrees with the existing literature in the Galilean case, but differs from most of it in the Carrollian case. Usually, generalized Carrollian limits are defined by sending $\omega \rightarrow \infty$ after performing the rescalings: 
\begin{align}
    \label{eq:oldCarr}
       X^{\tilde{A}} &= \frac{1}{\omega} x^{\tilde{A}} \,,& X^{A^\prime} &= x^{A^\prime} \nonumber\\
        E_{\mu}{}^{\tilde{A}} &= \frac{1}{\omega} e_{\mu}{}^{\tilde{A}} \,,  & E_{\mu}{}^{A^\prime} &= e_{\mu}{}^{A^\prime} \,, \nonumber \\
     \Lambda^{A^\prime}{}_{B^\prime} &= \lambda^{A^\prime}{}_{B^\prime} \,, & \Lambda^{\tilde{A}}{}_{\tilde{B}} &= \lambda^{\tilde{A}}{}_{\tilde{B}} \,,   & \Lambda^{A^\prime}{}_{\tilde{A}} &= \frac{1}{\omega} \lambda^{A^\prime}{}_{\tilde{A}} \,,
\end{align}
with time one of the $\tilde{A}$ directions and the $X^{A^\prime}$ spacelike. These rescalings are obtained by dividing the unified limit rescalings \eqref{eq:uniflim} by $\omega$. The limit that starts from the rescalings \eqref{eq:oldCarr} has been termed the ``$(D-p-2)$-brane Carroll limit" in the literature. The rationale for this name is that the $D-p-1$ rescaled $X^{\tilde{A}}$ coordinates are then viewed as directions longitudinal to the worldvolume of a Lorentzian $(D-p-2)$-brane and that $\omega \rightarrow \infty$ corresponds to sending $c$ to zero in the $p+1$ directions transverse to this brane. The results of the $(D-p-2)$-brane Carroll limit \eqref{eq:oldCarr} are, however, equivalent to those of the Euclidean $p$-brane Carroll limit \eqref{eq:uniflim}. In this paper, we will no longer use the term ``$(D-p-2)$-brane Carroll limit" and instead exclusively use ``Euclidean $p$-brane Carroll limit". In short, instead of letting time be a part of the brane worldvolume, as was commonly done, we now let the latter be determined by the coordinates that grow large in the limit. In this way, we wish to stress that viewing generalized Carrollian limits as Euclidean, as opposed to Lorentzian brane limits, can be more appropriate. The observation that the standard Carroll limit of a particle can be viewed as the Euclidean Carroll limit of a $(D-2)$ brane was also made in \cite{Ciambelli:2018xat,Petkou:2022bmz,Chabirand:2026sxe}. As mentioned in the introduction, this is natural from the perspective of timelike U-duality in string theory, as has already been noted in \cite{Blair:2023noj,Gomis:2023eav,Blair:2025nno}, where limits that are equivalent to our Euclidean Carroll limits have been obtained as time-like T-duals of generalized Galilean limits. Moreover, we will see in the next sections that the geometry that results from the Euclidean 0-brane Carroll limit naturally couples to a particle and not to a domain wall, as would be suggested by calling it a $(D-2)$-brane Carroll limit. The duality of non-Lorentzian limits provides another example showing that it is more suitable to view Carrollian limits as Euclidean brane limits. It was noticed in \cite{Barducci:2018wuj,Bergshoeff:2020xhv} that there is a duality that exchanges the results of a $p$-brane Galilei limit with those of a $(D-p-2)$-brane Carroll limit \eqref{eq:oldCarr}. This duality is performed by interpreting the longitudinal/transversal directions of the $p$-brane Galilei limit as the transversal/longitudinal ones of the $(D-p-2)$-brane Carroll limit and at the same time performing a Wick rotation in both the longitudinal and transversal directions. Reinterpreting the $(D-p-2)$-brane Carroll limit \eqref{eq:oldCarr} as a Euclidean $p$-brane Carroll limit \eqref{eq:uniflim}, makes the origin of this duality clear: both sides of the duality arise from the same limit, but with the time direction being viewed differently as a longitudinal or transversal one. This duality is summarized in Figure \ref{fig:diag1}.

This finishes our general discussion of how generalized Galilean and Carrollian limits can be dealt with in a unified way. In the rest of this paper, we will mainly focus on the cases in which there is only one longitudinal direction.

\section{Bargmann through the looking-glass: Alice} \label{sec:lookingglass}

In this section, we will first have a closer look at the four generalized Galilean and Carrollian limits whose definition involves one longitudinal or one transversal direction, and we will interpret them physically. Next, we will focus on the two limits with one longitudinal direction and argue that the non-Lorentzian symmetries that arise after taking them can be centrally extended.

According to the previous section, there are two generalized non-Lorentzian limits \eqref{eq:uniflim2} with $p=0$, i.e., with one longitudinal direction:
\begin{enumerate}
    \item the $0$-brane or ``particle" Galilei limit, which corresponds to the standard nonrelativistic limit,
    \item the Euclidean $0$-brane Carroll limit. 
\end{enumerate}
As discussed around \eqref{eq:Carrlimit}, the latter should not be confused with the usual Carroll limit, that, from the point of view of the unified limit \eqref{eq:uniflim2}, is the second of the following two limits with $p=D-2$, i.e., with one transversal direction: 
\begin{enumerate}
      \setcounter{enumi}{2}
    \item the $(D-2)$-brane Galilei limit, 
    \item the Euclidean $(D-2)$-brane Carroll limit, which is equivalent to the standard Carroll limit.
\end{enumerate}

\begin{figure}[t!] 
\centering
\begin{tikzpicture}
\node at (-4,0)
{
    \includegraphics[width=0.28\textwidth]{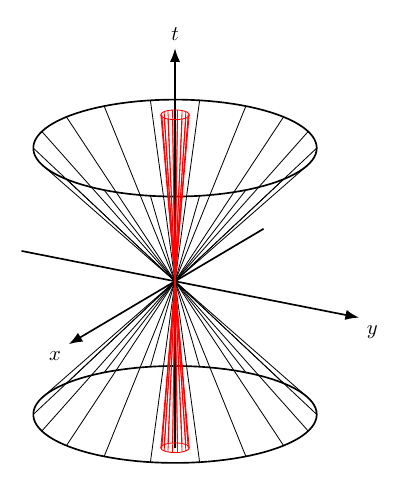}
};
\node at (3,0)
{
    \includegraphics[width=0.28\textwidth]{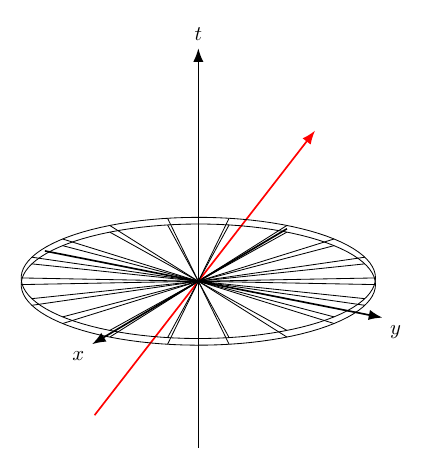}
};
\node at (-4,-6)
{    
      \includegraphics[width=0.28\textwidth]{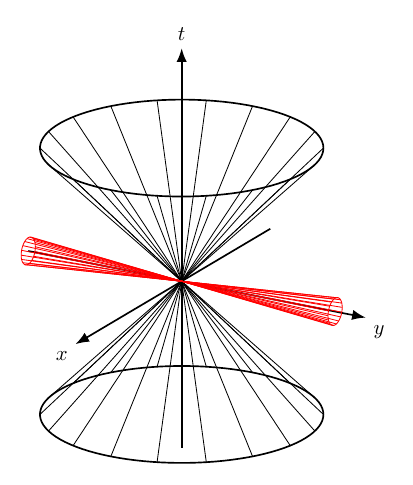}
};
\node at (3,-6)
{
    \includegraphics[width=0.28\textwidth]{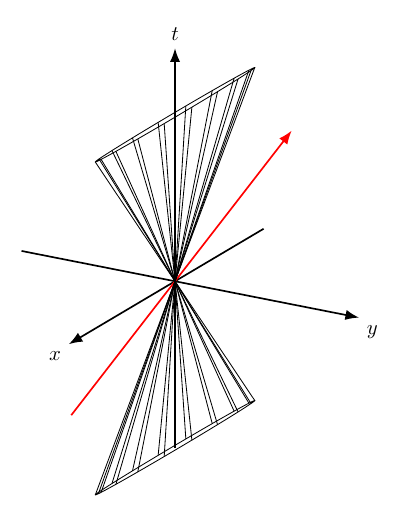}
};
\end{tikzpicture}
\caption{\label{fig:particlelimits}} Spacetime diagrams that illustrate the geometry of the $0$-brane Galilei (first row) and Euclidean $0$-brane Carroll (second row) limits. The limits zoom in on worldlines that lie inside the red narrow cones shown in the plots in the first column. The second column shows for each limit how the (black) lightcone is perceived from the perspective of a (red) worldline that the limit focuses on. 
\end{figure} 

\begin{figure}[t!] 
\centering
\begin{tikzpicture}
\node at (-4,0)
{
    \includegraphics[width=0.28\textwidth]{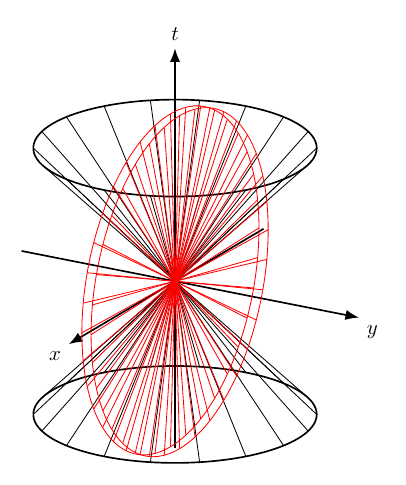}
};
\node at (3,0)
{
    \includegraphics[width=0.28\textwidth]{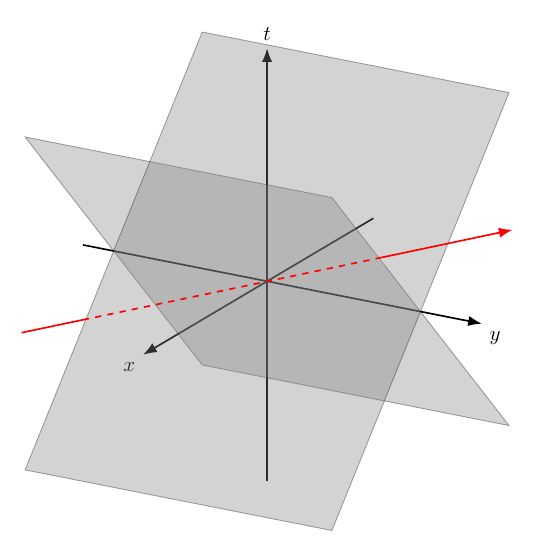}
};
\node at (-4,-6)
{    
      \includegraphics[width=0.28\textwidth]{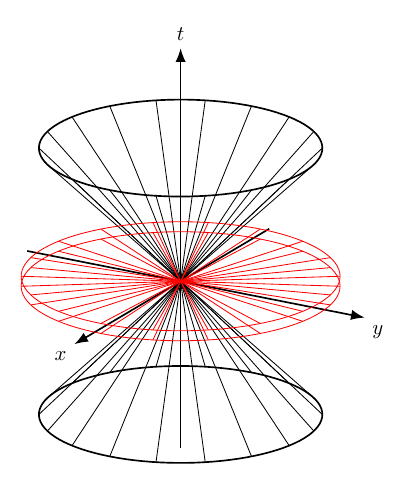}
};
\node at (3,-6)
{
    \includegraphics[width=0.28\textwidth]{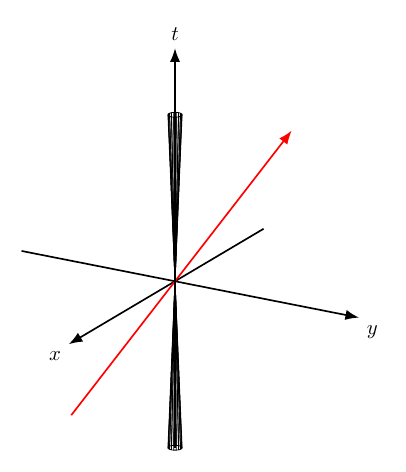}
};
\end{tikzpicture}
\caption{\label{fig:dwlimits}} Spacetime diagrams that illustrate the geometry of the $(D-2)$-brane Galilei (first row) and Euclidean $(D-2)$-brane Carroll (second row) limits. The limits zoom in on worldlines that lie outside the wide red cones shown in the plots in the first column. The second column shows for each limit how the (black) lightcone is perceived from the perspective of a (red) worldline that the limit focuses on. 
\end{figure}

Physically, the rescalings \eqref{eq:uniflim} that are used to define the above four limits have the effect of focusing on specific particle worldlines. This is illustrated in the spacetime diagrams of Figures \ref{fig:particlelimits} (for $p=0$) and \ref{fig:dwlimits} (for $p = D-2$). The first diagram on the top row of Figure \ref{fig:particlelimits} indicates that the $0$-brane Galilei limit zooms in on worldlines of physical particles that completely lie inside a narrow cone, whose axis coincides with the timelike direction. The $0$-brane Galilei limit thus focuses on bradyons\,\footnote{A bradyon is a particle that moves at speeds below the speed of light.} that move forward in time with very low speeds $v \ll c$. The Euclidean $0$-brane Carroll limit instead concentrates on tachyons that move forward with very large velocities $v_y \gg c$ in a particular spatial direction $y$. As illustrated in the first diagram on the second row of Figure \ref{fig:particlelimits}, this limit zooms in on worldlines that are situated inside a narrow cone around the $y$-axis. The Euclidean $(D-2)$-brane Carroll limit in Figure \ref{fig:dwlimits}, on the other hand, also zeroes in on tachyonic worldlines with speeds $v \gg c$. Unlike the Euclidean $0$-brane Carroll limit, these tachyons can, however, move in arbitrary spatial directions and are thus constrained to move outside of a very wide cone, whose axis is the time direction. This is shown in the first diagram on the second row of Figure \ref{fig:dwlimits}. Finally, the $(D-2)$-brane Galilei limit focuses on worldlines of bradyons and tachyons that have very small velocities $v_y \ll c$ in the spatial $y$-direction. These worldlines lie outside of a very wide cone whose axis coincides with the $y$-direction, and that is depicted in the first diagram in the first row of Figure \ref{fig:dwlimits}.

Note that in the strict $\omega \rightarrow \infty$ limit, the worldlines on which the $(D-2)$-brane Galilei and Euclidean $(D-2)$-brane Carroll limits zoom in, fill up the worldvolume of a Lorentzian and Euclidean domain wall respectively (in the 3-dimensional pictures of Figure \ref{fig:dwlimits} this is the worldsheet of a Lorentzian or Euclidean string). This gives a pictorial indication that these limits can be considered as adapted to domain walls, i.e., as the appropriate ones to take of $(D-2)$-brane actions. Likewise, the pictures of Figure \ref{fig:particlelimits} show that it is natural to take the $0$-brane Galilei and Euclidean $0$-brane Carroll limits of the particle actions of a bradyon (with timelike worldline) and tachyon (with spacelike worldline), respectively. We will elaborate more on this for $p=0$ in the next section. In what follows, we will concentrate on a different (albeit related) feature that distinguishes the two $p=0$ limits from their analogues with $p \neq 0$, namely that the spacetime symmetries that arise from them admit a central extension. For the $0$-brane Galilei limit, this corresponds to the fact that the ($0$-brane) Galilei algebra can be centrally extended to the Bargmann algebra. The similarity between the $0$-brane Galilei and Euclidean $0$-brane Carroll limits implies that the latter likewise gives rise to symmetries that can be centrally extended to what we will refer to as the `Alice algebra'. In the rest of this section, we will show how these Bargmann and Alice algebras can be obtained in a unified way from an In\"on\"u-Wigner contraction of the direct sum of the Poincar\'e algebra and an abelian generator.\,\footnote{It is also possible to obtain these algebras as subalgebras of the $(D+1)$-dimensional Poincar\'e algebra with one or two times, as the centralizer of a translation generator along a null direction. This is an algebraic version of the null reduction procedures that we will study in Section \ref{sec:nullred}.}

Before deriving the Bargmann and Alice algebras, let us first discuss the $p$-brane Galilei and Euclidean $p$-brane Carroll contractions of the Poincar\'e algebra. To do this in a unified way, we will find it convenient to make use of a sign parameter $\varepsilon = \mp 1$ to indicate whether time is a longitudinal or transversal direction and which limit or contraction we take. For a $p$-brane Galilei limit or contraction with time being longitudinal, we will have $\varepsilon = -1$, while for the Euclidean $p$-brane Carroll case with time in the transversal directions, we will have $\varepsilon = +1$. With this notation, we will denote the flat (Lorentzian or Euclidean) metrics that are used to contract $A^\prime$ and $\tilde{A}$ indices by $h_{A^\prime B^\prime}(\varepsilon)$ and $h_{\tilde{A} \tilde{B}}(\varepsilon)$, e.g., we have
\begin{align}
    \label{eq:epsnot}
    e_{\mu}{}^{A^\prime} e_{\nu A^\prime} \equiv e_{\mu}{}^{A^\prime} e_{\nu}{}^{B^\prime} h_{A^\prime B^\prime}(\varepsilon) \,,  \qquad \qquad  e_{\mu}{}^{\tilde{A}} e_{\nu \tilde{A}} \equiv e_{\mu}{}^{\tilde{A}} e_{\nu}{}^{\tilde{B}} h_{\tilde{A} \tilde{B}}(\varepsilon) \,.
\end{align}
Specifying $\varepsilon$ to be $\mp 1$, we have that
\begin{align}
    \label{eq:hsGal}
    &\varepsilon = -1 \ (\text{$p$-brane Galilei\ limit})\, :\hskip .5truecm h_{A^\prime B^\prime}(-1) = \eta_{A^\prime B^\prime} \qquad \text{and} \qquad h_{\tilde{A} \tilde{B}}(-1) = \delta_{\tilde{A}\tilde{B}} \,,
\end{align}
since time is a longitudinal direction, and that 
\begin{align}
    \label{eq:hsCarr}
    &\varepsilon = +1 \ (\text{Euclidean $p$-brane Carroll\ limit})\, :\hskip .5truecm h_{A^\prime B^\prime}(+1) = \delta_{A^\prime B^\prime} \quad \text{and} \quad h_{\tilde{A} \tilde{B}}(+1) = \eta_{\tilde{A}\tilde{B}} \,,
\end{align}
since time lies in the transversal directions.\,\footnote{We thus have that $\varepsilon = \mathrm{det}(h_{A^\prime B^\prime}(\varepsilon))$.} We then decompose the Poincar\'e generators in longitudinal translations $P_{A^\prime}$, transversal translations $P_{\tilde A}$, longitudinal rotations  $J_{A'B'}= -J_{B^\prime A^\prime}$, transversal rotations $J_{\tilde A \tilde B}= -J_{\tilde B \tilde A}$ and generalized boost transformations  $G_{A' \tilde A}$. In terms of this decomposition, the Poincar\'e algebra reads: 
\begin{align}
[G_{ A' \tilde A}\, , P_{\tilde B}] &=  h_{\tilde A \tilde B}(\varepsilon)\, P_{A'} \,, \qquad \qquad & [G_{ A' \tilde A}\, , P_{B'}] &= - h_{ A' B'}(\varepsilon)\, P_{\tilde A} \,, \nonumber \\ [G_{A^\prime \tilde{A}}\, , G_{B^\prime \tilde{B}}] &= - h_{A^\prime B^\prime}(\varepsilon) \, J_{\tilde{A}\tilde{B}} - h_{\tilde{A}\tilde{B}}(\varepsilon)\, J_{A^\prime B^\prime} \,, \nonumber 
\label{eq: Poincaré algebra}
\end{align}
Here, we have not given the trivial commutators and those involving the rotation generators $J_{A^\prime B^\prime}$ and $J_{\tilde A \tilde B} $ since these commutators just indicate how the generators transform as longitudinal or transversal (or mixed longitudinal-transversal) vectors or tensors, according to the indices they carry. 

Next, we consider the following In\"on\"u-Wigner redefinitions: 
\begin{equation}
J_{A^\prime B^\prime} = j_{A^\prime B^\prime} \,,\hskip .5truecm J_{\tilde A\tilde B} = j_{\tilde A\tilde B}\,,\hskip .5truecm    G_{A^\prime \tilde A} = \omega\, g_{A^\prime \tilde A}\,,  \hskip .5truecm P_{A^\prime} = \frac{1}{\omega} \, p_{A^\prime}\,,\hskip .5truecm P_{\tilde A} = p_{\tilde A}\,.
\end{equation}
Substituting these redefinitions into the Poincar\'e algebra and taking the limit that $\omega \to \infty$ we find the following non-zero commutator
\begin{equation}
[g_{A^\prime \tilde A} , p_{B^\prime}] = - h_{A^\prime B^\prime}(\varepsilon) \, p_{\tilde A}\,,
    \end{equation}
where again we have not given all commutators involving the longitudinal and transversal rotation generators. 

Let us now look at the case $p=0$ and denote the single longitudinal direction by $s$ ($A^\prime =s$), where $s$ stands for the time direction if $\varepsilon = -1$ and for a spatial direction if $\varepsilon = +1$. One can then have a central extension $m_s$ in the commutator of the $g_{s \tilde A}$ and $p_{\tilde B}$ generators.
This is possible because we have the invariant tensor $h_{\tilde A\tilde B}(\varepsilon)$ and because, for $p=0$, the $g_{s \tilde A}$ generator is a vector. This is clearly only true for $p=0$. For this reason, we expect (but do not prove!) that a similar central extension does not exist for $p\ne0$.\,\footnote{``Non-central extensions'' where the central charge generator carries a rotation index are possible. Furthermore, for $D=3$ the algebra allows a second central extension which we will not consider here. }
The existence of a central extension generator $m_s$ for $p=0$ can either be proven by checking the relevant Jacobi identities containing $m_s$ or by extending the Poincar\'e algebra with an abelian generator $M$ that commutes with all the other generators.\,\footnote{The Noether symmetry corresponding to this abelian generator corresponds to the conservation of the difference in the number of particles and anti-particles.} In the latter case one should use the following modified  In\"on\"u-Wigner redefinitions: 
\begin{align}
     J_{\tilde A \tilde B}&=j_{\tilde A \tilde B}\,, &&  P_{s}= \frac{1}{2 \omega}\, p_s + \omega \, m_s \,, && M= \frac{1}{2 \omega} \, p_{s} - \omega\, m_s  \,, \nonumber\\[1ex]  G_{s\tilde A}&= \omega\, g_{s \tilde A}\,, && P_{\tilde A}= p_{\tilde A}\,.
     \label{eq: extended Inonu-Wigner contraction}
\end{align}
In both cases, one finds the following non-zero commutators that define the Bargmann and Alice algebras: 
\begin{alignat}{3}
    \label{Bargmann}
 &   {\rm \bf Bargmann\ algebra\,:}\qquad \ \  &
 [g_{s\tilde A}, p_s] &= p_{\tilde A}\,,\hskip .8truecm &
 [g_{s\tilde A}, p_{\tilde B} ] &= \delta_{\tilde A\tilde B}\, m
  \,,\cr
&{\rm \bf Alice\ algebra\,:}\qquad \ \ \hskip 1truecm &
 [g_{s\tilde A}, p_s] &= -p_{\tilde A}\,,\hskip .8truecm &
 [g_{s\tilde A}, p_{\tilde B} ] &= \eta_{\tilde A\tilde B}\, p
  \,,
\end{alignat}
where we have renamed $m_s$ to $m$ in the Bargmann algebra and to $p$ in the Alice algebra, in anticipation of their interpretation in the particle actions that we will discuss in the next section. Like before, we have refrained from giving the explicit commutators that involve transversal rotation (for Bargmann) or Lorentz (for Alice) generators. 

Given these centrally extended Bargmann and Alice algebras, it is interesting to find systems in which the central charge is non-trivially realized. Let us now turn to showing that there exist particle actions that give examples of such systems. 

\section{Critical Bargmann and Alice Particles} \label{sec:BargmannAlice}

Here, we will show that the centrally extended Bargmann and Alice algebras arise naturally as symmetry algebras of the $0$-brane Galilei and Euclidean $0$-brane Carroll limits of the action of a bradyon and a tachyon, coupled to an appropriately tuned one-form gauge potential. The latter will be done using the unified approach presented in Section \ref{sec:unification}, which allows us to take these limits simultaneously. In the case of the $0$-brane Galilei limit, this leads to the Bargmann particle action, i.e., the action of a massive nonrelativistic particle. For the Euclidean $0$-brane Carroll limit, one obtains the action of a novel type of particle, that we will call the ``Alice particle". We will also see how the Bargmann and Alice particle actions can be modified to incorporate the effect of a cosmological constant and for which signs of the cosmological constant these particles undergo stable motion. Since the limits discussed in this section involve the introduction of a fine-tuned gauge potential, they are usually called ``critical". There also exists literature on (classical and quantum) Galilei and Carroll (i.e., Euclidean $(D-2)$-brane Carroll) particles that are not critical limits of relativistic particles (e.g., \cite{Souriau1997,Duval:2014uoa,Bergshoeff:2014jla,Bergshoeff:2015wma,deBoer:2021jej,Bergshoeff:2022eog,Henneaux:2021yzg,Ecker:2024czx,Figueroa-OFarrill:2023vbj,Figueroa-OFarrill:2023qty,Figueroa-OFarrill:2024ocf}). These do not have non-zero values of a central charge, since the usual Carroll algebra can (in generic dimensions) not be centrally extended and only critical Galilei limits lead to a non-trivial central charge.

To explain the notion of a critical limit of a particle,\,\footnote{Critical limits can in generic dimensions also be taken for Lorentzian or Euclidean $p$-branes with $p>0$, by coupling to a $(p+1)$-form potential to cancel the brane's rest tension. Whereas for $p=0$, critical limits lead to central extensions, for $p>0$ they typically lead to non-central extensions in the resulting non-Lorentzian spacetime symmetry algebras. In particular, one can take a critical usual Carroll limit, interpreted as a Euclidean $(D-2)$-brane Carroll limit, of a Euclidean domain wall action. Thermodynamic properties of the resulting objects have been studied in \cite{Ciambelli:2018xat,Petkou:2022bmz,Chabirand:2026sxe}. As discussed below \eqref{eq:oldCarr}, the fact that the Euclidean $0$-brane and Euclidean $(D-2)$-brane Carroll limits can be applied as critical limits of a particle and domain wall action, respectively, gives one justification to no longer use the terms $(D-2)$-brane and $0$-brane Carroll limits (without the adjective ``Euclidean"), that can be found in the literature.} let us first look again at the nonrelativistic ($0$-brane Galilei) limit, illustrated in the top left picture of Figure \ref{fig:particlelimits}. Alternatively, one can say that this limit focuses on the worldlines of bradyons whose energy is dominated by the rest energy. A critical limit is one in which this rest energy contribution is cancelled. This can be done by coupling the particle to a background one-form gauge potential and tuning its timelike component to precisely cancel the particle's rest energy. The nonrelativistic approximation to the dispersion relation and worldline action of a relativistic particle then becomes exact. Note that the introduction of a background one-form gauge field introduces an extra symmetry that, via Noether's theorem, corresponds to conservation of the difference in the number of particles and anti-particles. The critical limit decouples the anti-particles and, therefore, leaves one with a conserved particle number (or equivalently mass) charge. It is the latter charge that centrally extends the Galilei algebra to the Bargmann algebra, thus showing that the critical limit gives a nonrelativistic particle action in which the central charge is non-trivially realized. Since the Euclidean $0$-brane Carroll limit is formally the same as the $0$-brane Galilei one, with the time direction exchanged for a particular spatial direction $y$, one can also state that it concentrates on tachyons, for which the dominant contribution to the momentum in the $y$-direction is given by a rest momentum. It is then suggestive that one can take a critical Euclidean $0$-brane Carroll limit of a relativistic tachyon, coupled to a one-form gauge potential to cancel the tachyon's rest momentum in the $y$-direction. The resulting particle has a non-zero value of the central charge of the Alice algebra and can be called ``the Alice particle".

To make the previous paragraph concrete, we simultaneously take the $0$-brane Galilei and Euclidean $0$-brane Carroll limits of the following relativistic massive particle action, coupled to a one-form gauge potential $M_{\mu}$:
\begin{equation}
\label{S}
    S = \varepsilon M c \int \rmd \lambda\, \left[\sqrt{\varepsilon G_{\mu\nu} \dot{X}^\mu \dot{X}^\nu} -  M_{\mu}{\dot X}^{\mu}\right] = \varepsilon M c\int \rmd\lambda\,\left[\sqrt{\varepsilon {\dot X}^{A} {\dot X}_{A}} -  M_{A}{\dot X}^{A}\right]\,.
    \end{equation}
Here, we have used the sign parameter $\varepsilon$, introduced around eqs. \eqref{eq:epsnot}, \eqref{eq:hsGal} and \eqref{eq:hsCarr}, $M$ is a parameter with the dimension of mass, $\lambda$ parametrizes the worldline and a dot denotes a derivative with respect to $\lambda$ ($\dot{X}^{\mu} \equiv \tfrac{\rmd X^\mu}{\rmd \lambda}$). In the second equality, we have written the metric $G_{\mu\nu}$ in terms of a Vielbein $E_{\mu}{}^{A}$ (with inverse $E^{\mu}{}_{A}$) as in \eqref{eq:metricvielb} and we have defined:
\begin{equation}
{\dot X}^{A} \equiv {\dot X}^\mu \, E_\mu{}^{A}\,,\hskip 1.5truecm 
M_{A} \equiv E^{\mu}{}_{A}\,  M_{\mu} \,.\end{equation}
For $\varepsilon=-1$, in which case we will take the $0$-brane Galilei limit, the action \eqref{S} is that of a bradyon with ${\dot X}^{A} {\dot X}_{A} <0$. When we consider the Euclidean $0$-brane Carroll limit and thus take $\varepsilon=+1$, the same action describes a tachyon with ${\dot X}^{A} {\dot X}_{A} > 0$. Note that in \eqref{S}, we have tuned the coupling constant in front of the gauge potential $M_{\mu}$ to be equal to the mass. This fine-tuning is an important part of the critical limit.

Since we focus on a particle limit (with $p=0$), we only have one longitudinal direction.\,\footnote{It is this longitudinal direction that is boost invariant after taking the limit. Note that, in contrast to the standard Carroll case, the Euclidean $0$-brane Carroll limit leads to only one spatial boost-invariant direction.} As in the In\"on\"u-Wigner contraction \eqref{eq: extended Inonu-Wigner contraction}, we decompose
$A =\big(s, \tilde A\big)$
where $s$ stands for the single longitudinal direction that can be time or a spatial direction. We next make the following redefinitions including the vector field $M_\mu$ and the mass parameter $M$:
\begin{equation}\label{redef11}
E_\mu{}^s = \omega\, e_\mu{}^s\,,\hskip 1 truecm  E_\mu{}^{\tilde A} = e_\mu{}^{\tilde A}\,,\hskip 1truecm 
 M_\mu = \omega\, e_\mu{}^s + \omega^{-1}\,m_\mu\,,\hskip 1truecm M=\omega\, m \,.
\end{equation}
Substituting these redefinitions into the action \eqref{S} and expanding the square root, we obtain 
\begin{equation}
S = \varepsilon\, m c\,\omega\int \rmd\lambda
\left[\omega {\dot X}^s +\frac{1}{2}\frac{\varepsilon{\dot X}^{\tilde A}{\dot X}_{\tilde A}}{\omega{\dot X}^{s}}
-\omega {\dot X}^s -\omega^{-1}m_\mu{\dot X}^\mu + \mathcal{O}(\omega^{-3}) \right] \,,
\end{equation}
where
\begin{equation}
\label{eq:XsXAt}
{\dot X}^s \equiv {\dot X}^\mu e_\mu{}^{s}\,,\hskip 1truecm 
{\dot X}^{\tilde A} \equiv {\dot X}^\mu e_\mu{}^{\tilde A}\,, \hskip 1truecm \dot{X}^{\tilde{A}} \dot{X}_{\tilde{A}} \equiv \dot{X}^{\tilde{A}} \dot{X}^{\tilde{B}} h_{\tilde{A}\tilde{B}}(\varepsilon) \,.
\end{equation}
We see that the $\omega^2$ terms cancel so that, after taking the $\omega\to \infty$ limit, we are left with the following particle action:
\begin{equation}
\label{eq:BargmannAliceunified}
S = m c\,\int \rmd\lambda
\left[\frac{1}{2}\frac{{\dot X}^{\tilde A}{\dot X}_{\tilde A}}{{\dot X}^{s}}
 - \varepsilon\, m_\mu{\dot X}^\mu\right]\,.
\end{equation}

To show the nature of this particle in the Bargmann case with $\varepsilon = -1$, we consider flat Minkowski spacetime with $X^\mu=\big(X^s = ct,X^{\tilde A}\big)$ and take $m_\mu=(-\phi/ c^2, -\phi_{\tilde{A}}/c)$, to obtain 
\begin{align}
S &= m c\,\int \rmd\lambda\, \left[
\frac{1}{2} \frac{
\frac{\rmd X^{\tilde A}}{\rmd\lambda}
\frac{\rmd X^{\tilde B}}{\rmd\lambda}\delta_{\tilde A\tilde B}}{
\frac{\rmd (ct)}{\rmd\lambda}} - \frac{\phi}{c} \frac{\rmd t}{\rmd \lambda} - \frac{\phi_{\tilde{A}}}{c} \frac{\rmd X^{\tilde{A}}}{\rmd \lambda} \right] \nonumber \\
&= m \,\int \rmd t\, \left[ \frac{1}{2}
\frac{\rmd X^{\tilde A}}{\rmd t}
\frac{\rmd X^{\tilde B}}{\rmd t}\delta_{{\tilde A}{\tilde B}} - \phi - \phi_{\tilde{A}} \frac{\rmd X^{\tilde{A}}}{\rmd t} \right] \,.
\end{align}
This is the standard action of a nonrelativistic particle of mass $m$ that is subjected to the Newton potential $\phi$ and velocity-dependent fictitious forces, captured by the $\phi_{\tilde{A}} \rmd X^{\tilde{A}}/\rmd t$ term. Note that the Newton potential $\phi$ has dimensions of (length/time)${}^2$ here. By taking $\phi$ and $\phi_{\tilde{A}}$ to be zero, we get the action of a free nonrelativistic massive particle, the Bargmann particle:
\begin{align}
\label{eq:Bargmann}
    S_{\text{Bargmann}} = \frac{m}{2} \int \rmd t \, \frac{\rmd X^{\tilde A}}{\rmd t}
\frac{\rmd X^{\tilde B}}{\rmd t}\delta_{{\tilde A}{\tilde B}} \,.
\end{align}
Similarly, for the Carrollian case $\varepsilon = +1$, we obtain the following action in flat spacetime with coordinates $X^\mu = (X^{\tilde{A}}, X^s = y)$ and $m_\mu = (\psi_{\tilde{A}},\psi)$:
\begin{equation}
\label{eq:Aliceinteracting}
S  
= p \,\int \rmd y\, \left[ \frac{1}{2}
\frac{\rmd X^{\tilde A}}{\rmd y}
\frac{\rmd X^{\tilde B}}{\rmd y}\eta_{{\tilde A}{\tilde B}} - \psi - \psi_{\tilde{A}} \frac{\rmd X^{\tilde{A}}}{\rmd y} \right] \,,
\end{equation}
where we have defined $p \equiv m c > 0$, which has the dimension of a (positive) momentum, as opposed to the positive mass of the Bargmann particle. In \eqref{eq:Aliceinteracting}, the role of the worldline parameter is played by $y$\,\footnote{Note that this is the coordinate that is invariant under the Carrollian boosts, so it can sensibly act as a worldline parameter. Additionally, as we show in Appendix \ref{sec:disprels}, a free Alice particle can only move either forward or backward in $y$. The $y$-direction thus shares some similarities with time as an evolution parameter.} and the potential $\psi$ is now dimensionless. The free version of \eqref{eq:Aliceinteracting}, in which $\psi = 0 = \psi_{\tilde{A}}$, is what we will call the action of the Alice particle:
\begin{equation}
\label{eq:Alice}
S_{\text{Alice}} 
= \frac{p}{2}\,\int \rmd y\, 
\frac{\rmd X^{\tilde A}}{\rmd y}
\frac{\rmd X^{\tilde B}}{\rmd y}\eta_{{\tilde A}{\tilde B}} \,.
\end{equation}
Note that critical decoupling limits of actions for spacelike branes in flat space have been discussed in \cite{Gomis:2023eav}. The above Alice particle action then corresponds to the special case of the limit of a zero-dimensional spacelike brane action.

Upon quantizing, one finds that wave functions that describe a  Bargmann particle \eqref{eq:Bargmann} obey the Schr\"odinger equation. Likewise, quantization of the Alice particle \eqref{eq:Alice} leads to what we will refer to as the Euclidean Carroll-Schr\"odinger equation. The latter was studied in two spacetime dimensions in \cite{Najafizadeh:2024imn,Rojas:2025rot,Rojas:2025ygg} (where it was called the Carroll-Schr\"odinger equation). In Appendix \ref{sec:SchrCarrSchr}, we show how the Schr\"odinger and Euclidean Carroll-Schr\"odinger equations in arbitrary spacetime dimensions can be derived in our unified approach as critical limits of a relativistic massive complex scalar theory. The dispersion relations of the Schr\"odinger and Euclidean Carroll-Schr\"odinger equations (and thus of the Bargmann and Alice particles) can analogously be obtained via a unified critical limit, as we show in Appendix \ref{sec:disprels}. This shows that the critical limits that lead to the Bargmann and Alice particles decouple part of the spectrum of states of the relativistic particle that one starts from. 

The Bargmann and Alice actions give Lagrangians that are only boost invariant up to a total derivative, and this leads to modified Noether charges. It is these modifications that give rise to the central extensions of the Bargmann and Alice algebras \eqref{Bargmann} that we encountered in the previous section. Indeed, applying Noether's theorem to the Bargmann and Alice actions, one gets the following conserved charges for the boost transformations: 
\begin{align}
  \text{Bargmann: } \qquad  g_{s \tilde{A}} = - m\,  t\, \frac{\rmd X_{\tilde{A}}}{\rmd t} +  m\, X_{\tilde{A}} \,, \qquad \qquad \quad \text{Alice: } \qquad   g_{s \tilde{A}} = -p\, y \, \frac{\rmd X_{\tilde{A}}}{\rmd y} + p\, X_{\tilde{A}} \,.
\end{align}
Here, the last term proportional to $X_{\tilde{A}}$ corresponds to the total derivative to which the Bargmann and Alice Lagrangians vary under boosts. The conserved charges associated to translations in the $X^{\tilde{A}}$ directions are given by 
\begin{align}
    \label{eq:momenta}
    \text{Bargmann: } \qquad p_{\tilde{A}} = m\, \frac{\rmd X_{\tilde{A}}}{\rmd t} \,, \qquad \qquad \qquad \text{Alice: } \qquad p_{\tilde{A}} = p\, \frac{\rmd X_{\tilde{A}}}{\rmd y} \,.
\end{align}
Using the canonical Poisson brackets between the transversal coordinates and momenta
\begin{align}
    \label{eq:PB}
    \left\{X^{\tilde{A}}, X^{\tilde{B}}\right\} = 0 \,, \qquad \qquad \left\{p_{\tilde{A}}, p_{\tilde{B}}\right\} = 0 \,, \qquad \qquad \left\{X^{\tilde{A}}, p_{\tilde{B}}\right\} = \delta_{\tilde{B}}^{\tilde{A}} \,,
\end{align}
as well as that $X_{\tilde{A}} = \delta_{\tilde{A}\tilde{B}} X^{\tilde{B}}$ in the Bargmann case and $X_{\tilde{A}} = \eta_{\tilde{A}\tilde{B}} X^{\tilde{B}}$ in the Alice case, one then finds the following Poisson brackets of the boost and transversal translation charges: 
\begin{align}
    \text{Bargmann: } \qquad \left\{g_{s\tilde{A}}, p_{\tilde{B}}\right\} = \delta_{\tilde{A}\tilde{B}}\, m \,, \qquad \qquad \qquad \text{Alice: } \qquad \left\{g_{s\tilde{A}}, p_{\tilde{B}}\right\} = \eta_{\tilde{A}\tilde{B}}\, p \,.
\end{align}
The Bargmann and Alice algebras \eqref{Bargmann} are thus indeed realized as Poisson bracket algebras.

The flat space Bargmann and Alice particle actions \eqref{eq:Bargmann} and \eqref{eq:Alice} can be generalized to include the effect of a cosmological constant. For the Bargmann particle, this entails generalizing the underlying Bargmann symmetries to the so-called (centrally extended) Newton-Hooke symmetries \cite{Bacry:1968zf,Derome:1972,Dubois:1973,Brugues:2006yd,Gibbons:2003rv}, whose algebra is obtained as an In\"on\"u-Wigner contraction of a direct sum of the anti-de Sitter (adS) or de Sitter (dS) algebras with an extra abelian generator. To derive these generalizations from a limit in our unified approach, we introduce a set of local coordinates $X^{\mu} = (S_1,\, S_2,\, U^i)$ and start from \eqref{S} in the background metric and gauge field\,\footnote{In the following, we set $c=1$ for simplicity.}
\begin{align}
    \label{eq:staticpatch}
    & \rmd s^2 = G_{\mu\nu} \rmd X^\mu  \rmd X^\nu = \varepsilon  \left(1 + \varepsilon \Lambda_0 \,S_1{}^2 \right) (\rmd S_2)^{2} - \varepsilon \frac{(\rmd S_1)^2}{1 + \varepsilon \Lambda_0 \,S_1{}^2 } + S_1{}^2 g_{ij}(U,\varepsilon) \rmd U^i \rmd U^j
    \,, \nonumber \\  & M_\mu\,  \rmd X^\mu = \rmd S_2  \,.
\end{align}
Here, the metric $g_{ij}(U,\varepsilon) \rmd U^i \rmd U^j$ is either the metric on a sphere or a hyperboloid,\,\footnote{This hyperboloid is defined by the equation $\eta_{\alpha \beta} y^{\alpha} y^{\beta}= -1$, embedded in $\mathbb{R}^{1,D-2}$, and the metric $g_{ij}(U,1) \rmd U^i \rmd U^j$ is the one induced from the flat metric $\eta_{\alpha\beta} \rmd y^\alpha \rmd y^\beta$. Here, the index $\alpha$ assumes $D-1$ values and $\eta_{\alpha \beta}$ is the Minkowski metric with mostly plus signature.} depending on the value of $\varepsilon$:
\begin{align}
    \varepsilon = -1 \ &: \ \ \ g_{ij}(U,-1) \rmd U^i \rmd U^j = \text{metric on a $(D-2)$-dimensional sphere of unit radius} \,, \nonumber \\
    \varepsilon = +1 \ &: \ \ \ g_{ij}(U,1) \rmd U^i \rmd U^j = \text{metric on a $(D-2)$-dimensional hyperboloid} \,.
\end{align}
For both possible values of $\varepsilon$, \eqref{eq:staticpatch} is the metric on (part of) adS spacetime when $\Lambda_0 < 0$ and on (part of) dS spacetime when $\Lambda_0 > 0$. In all cases, $\Lambda_0$ is related to the true cosmological constant $\Lambda$ by a dimension-dependent positive factor. 
For $\varepsilon = -1$, the metric of \eqref{eq:staticpatch} corresponds to either the metric of adS spacetime in Schwarzschild coordinates or the metric on the static patch of dS spacetime. When $\varepsilon = +1$ and $\Lambda_0 > 0$, one obtains the dS metric in a patch foliated by slices of a product of $\mathbb{R}$ and a $(D-2)$-dimensional hyperboloid \cite{Anninos:2012qw}. 

Upon using the following rescalings
\begin{align}
    S_2=\omega \, s_2\,, \qquad S_1=s_1\,, \qquad \Lambda_0=\frac{K}{\omega^2}\,, \qquad U^i=u^i\,, \qquad M= \omega\, m_s \,,
\end{align}
in the action \eqref{S}, one finds that the $\omega \rightarrow \infty$ limit is well-defined and leads to the following action:
\begin{align}
    \label{eq:unifNH}
    S &= \frac{m_s}{2}\int \rmd s_2  \left[- \varepsilon \bigg(\frac{\rmd s_{1}{}}{\rmd s_{2}{}}\bigg)^2 +  s_1{}^2 g_{ij}(u,\varepsilon) \frac{\rmd u^i}{\rmd s_2}\frac{\rmd u^j}{\rmd s_2} +  K s_1{}^2 \right] \,.
\end{align}
Let us now discuss the two $\varepsilon = \mp 1$ cases in turn. For $\varepsilon = -1$, we rename 
\begin{align}
    s_1 = r \,, \qquad s_2 = t \,, \qquad u^i = \theta^i \,, \qquad g_{ij}(u,-1) = g_{ij}(\theta) \,, \qquad m_s = m \,,
\end{align}
to allow for an easier comparison with the Bargmann particle action. Indeed, the action \eqref{eq:unifNH} then becomes
\begin{align}
    S_{\text{Bargmann\,,}\, \Lambda} &= \frac{m}{2} \int \rmd t \, \left(\left(\frac{\rmd r}{\rmd t}\right)^2 + r^2 g_{ij}(\theta) \frac{\rmd \theta^i}{\rmd t} \frac{\rmd \theta^j}{\rmd t} + K  r^2 \right)  \,.
\end{align}
This action describes the motion of a nonrelativistic Bargmann particle, under the influence of a quadratic potential that arises from the cosmological constant, in spherical coordinates. Reverting back to Cartesian coordinates, it can be rewritten as \cite{Gibbons:2003rv,Brugues:2006yd}:
\begin{align}
    \label{eq:BargmannLambda}
    S_{\text{Bargmann\,,}\, \Lambda} &= \frac{m}{2} \int \rmd t \, \left(\frac{\rmd X^{\tilde{A}}}{\rmd t} \frac{\rmd X^{\tilde{B}}}{\rmd t} \delta_{\tilde{A}\tilde{B}} + K \,  X^{\tilde{A}} X^{\tilde{B}} \delta_{\tilde{A}\tilde{B}} \right)  \,.
\end{align}
The motion of this particle is either oscillatory or exponentially growing or decaying. Stable, oscillatory behaviour is obtained when $K < 0$, i.e., when the background metric \eqref{eq:staticpatch} (for $\varepsilon = -1$) is the adS metric. When instead $K > 0$ and the background metric \eqref{eq:staticpatch} (with $\varepsilon = -1$) is dS, the particle exhibits unstable, exponential runaway motion. Note that the action \eqref{eq:BargmannLambda} for the Bargmann particle with a cosmological constant can also be obtained as a nonlinear realization of the centrally extended Newton-Hooke symmetry algebra (see e.g., \cite{Bergshoeff:2022eog}).

For $\varepsilon = + 1$, the action \eqref{eq:unifNH} is given by
\begin{align}
    \label{eq:AliceLambda1}
    S_{\text{Alice\,,}\, \Lambda} &= \frac{p}{2} \int \rmd y \, \left(- \left(\frac{\rmd \tau}{\rmd y}\right)^2 + \tau^2 h_{ij}(u) \frac{\rmd u^i}{\rmd y} \frac{\rmd u^j}{\rmd y} + K \tau^2 \right)  \,,
\end{align}
where we have renamed
\begin{align}
    s_1 = \tau \,, \qquad s_2 = y \,, \qquad g_{ij}(u,1) = h_{ij}(u) \,, \qquad m_s = p \,.
\end{align}
Note that $-\rmd \tau^2 + h_{ij} \rmd u^i \rmd u^j$ is the metric of Minkowski spacetime in hyperbolic coordinates. In standard inertial Minkowski coordinates $X^{\tilde{A}}$, for which $\tau^2 = - \eta_{\tilde{A}\tilde{B}} X^{\tilde{A}} X^{\tilde{B}}$, the action \eqref{eq:AliceLambda1} can be rewritten as
\begin{align}
    \label{eq:AliceLambda}
    S_{\text{Alice\,,}\, \Lambda} &= \frac{p}{2} \int \rmd y \, \left( \frac{\rmd X^{\tilde{A}}}{\rmd y} \frac{\rmd X^{\tilde{B}}} {\rmd y} \eta_{\tilde{A} \tilde{B}} - K \, X^{\tilde{A}} X^{\tilde{B}} \eta_{\tilde{A}\tilde{B}} \right) \,.
\end{align}
As for the Bargmann particle, the cosmological constant gives rise to a quadratic potential, in which the Alice particle moves. In this case, the motion of the particle is oscillatory and stable when $K > 0$, i.e., when the background metric \eqref{eq:staticpatch}, with $\varepsilon = +1$, is dS. When $K < 0$ and \eqref{eq:staticpatch} with $\varepsilon = +1$ is the adS metric, the Alice particle exhibits unstable, exponential runaway behaviour. Although we have not yet checked it, we expect that there is a generalization of the Alice algebra that includes a cosmological constant and from which the action \eqref{eq:AliceLambda} can be obtained as a nonlinear realization, in analogy to what happens in the Bargmann case \eqref{eq:BargmannLambda}.

After having obtained the Bargmann and Alice particle actions via a non-Lorentzian limit, we now turn to deriving them in a different manner, namely via null reduction. 

\section{Bargmann and Alice from Null Reduction} \label{sec:nullred}

A theory with Bargmann symmetries in $D$ spacetime dimensions can be conveniently obtained from a spacelike null reduction of a  relativistic theory in $D+1$ spacetime dimensions with signature 
$(1,D)$ \cite{Duval:1984cj}. Even though one reduces over a lightlike direction, we will call this null reduction spacelike, since it can be viewed as a limit of reductions over spacelike circles \cite{Seiberg:1997ad,Sen:1997we}. Here, we will show that one can similarly construct a theory with Alice symmetries in $D$ spacetime dimensions, provided that one performs a timelike null reduction of a relativistic theory in $D+1$ spacetime dimensions with signature $(2,D-1)$. We refer to this reduction as timelike, since it can be defined as a limit of reductions over timelike circles. In this section, we will first derive the metric Ansatz for both the spacelike and timelike null reductions by taking a limit of the Ans\"atze of reductions over spacelike and timelike circles in a unified manner.  We will then proceed to show how the Bargmann and Alice particle actions can be obtained from spacelike and timelike null reductions of the massless particle action in $(D+1)$-dimensional spacetimes, with one and two times respectively.

Let us start from a $(D+1)$-dimensional spacetime with local coordinates $X^M = (X^\mu, X^{s_1})$, where $X^\mu$ are $D$ noncompact coordinates and $X^{s_1}$ parametrizes a (spacelike or timelike) circle. We then consider the following Ansatz for the reduction of a $(D+1)$-dimensional metric $\hat{G}_{MN}$ to $D$ dimensions:
\begin{align}
    \label{eq:standardKK}
    \rmd \hat{s}^2 &= \hat{G}_{MN} \rmd X^M \rmd X^N = - \varepsilon\, \rme^{2 \beta \Phi} \left(\rmd X^{s_1} + M_\mu \rmd X^\mu\right)^2 + G_{\mu\nu} \rmd X^\mu \rmd X^\nu \nonumber \\
    &= - \varepsilon\, \rme^{2 \beta \Phi} \left(\rmd X^{s_1} + M_\mu \rmd X^\mu\right)^2 + \varepsilon E_{\mu}{}^{s_2} E_{\nu}{}^{s_2} \rmd X^\mu \rmd X^\nu + E_{\mu}{}^{\tilde{A}} E_{\nu}{}^{\tilde{B}} h_{\tilde{A}\tilde{B}}(\varepsilon) \rmd X^\mu \rmd X^\nu \,.
\end{align}
In the second line, we have written the $D$-dimensional metric $G_{\mu\nu}$ in terms of a Vielbein $E_{\mu}{}^{A}$ and decomposed its flat index $A$ as $A = \{s_2, \tilde{A}\}$.\,\footnote{Note that $s_2$ is a flat index, while $s_1$ is a curved one.} We have also used the sign parameter $\varepsilon = \mp 1$ and flat metric $h_{\tilde{A}\tilde{B}}(\varepsilon)$ introduced around eqs. \eqref{eq:epsnot}, \eqref{eq:hsGal} and \eqref{eq:hsCarr}. As usual for a reduction Ansatz, it is assumed that the $D$-dimensional fields $E_{\mu}{}^{s_2}$, $E_{\mu}{}^{\tilde{A}}$ and $M_\mu$ do not depend on the compact coordinate $X^{s_1}$. Finally, $\beta$ in \eqref{eq:standardKK} is a non-zero real number, for which we do not specify a particular value as it will not be important in what follows. 

In case $\varepsilon = -1$, the $(D+1)$-dimensional metric of the Ansatz \eqref{eq:standardKK} has signature $(1,D)$, with $s_2$ corresponding to the timelike direction. The compact direction $X^{s_1}$ is spacelike, so that \eqref{eq:standardKK} is recognized as the standard Kaluza-Klein Ansatz for a spacelike reduction. By contrast, when $\varepsilon = +1$, the $(D+1)$-dimensional metric \eqref{eq:standardKK} has two time directions, one contained in the $\tilde{A}$ directions and the other corresponding to the compact direction $X^{s_1}$. For $\varepsilon = +1$, \eqref{eq:standardKK} is thus the appropriate Ansatz for a timelike reduction of a theory with two times to a lower-dimensional one with a single time.

Next, we perform the following redefinitions: 
\begin{align}
\label{eq:KKredefs}
    E_{\mu}{}^{s_2} &= \omega \, e_{\mu}{}^{s_2} \,, \qquad  E_{\mu}{}^{\tilde{A}} = e_{\mu}{}^{\tilde{A}} \,, \qquad  M_{\mu} = \omega^2 \, \rme^{-\beta \phi}\, e_{\mu}{}^{s_2} +  \, m_{\mu} \,, \qquad  \rme^{\beta \Phi} = \frac{\rme^{\beta \phi}}{\omega} \,.
\end{align}
to turn \eqref{eq:standardKK} into:
\begin{align}
    \label{eq:rescstandardKK}
    \rmd \hat{s}^2 &= - \varepsilon \frac{\rme^{2\beta \phi}}{\omega^2} \left(\rmd X^{s_1} + \omega^2 \rme^{-\beta \phi} e_{\mu}{}^{s_2} \rmd X^\mu + m_{\mu} \rmd X^{\mu} \right)^2 + \varepsilon \omega^2 e_{\mu}{}^{s_2} e_{\nu}{}^{s_2} \rmd X^\mu \rmd X^\nu  \nonumber \\ & \qquad + e_{\mu}{}^{\tilde{A}} e_{\nu}{}^{\tilde{B}} h_{\tilde{A} \tilde{B}}(\varepsilon) \rmd X^\mu \rmd X^\nu \,. 
\end{align}
Expanding the square in the first term, one finds two terms of the form $\varepsilon \omega^2 e_{\mu}{}^{s_2} e_{\nu}{}^{s_2} \rmd X^\mu \rmd X^\nu$ that cancel. The $\omega \rightarrow \infty$ limit of \eqref{eq:rescstandardKK} is then well-defined and yields:
\begin{align}
    \rmd \hat{s}^2 &= - 2 \varepsilon \rme^{\beta \phi} e_{\mu}{}^{s_2} \rmd X^\mu \rmd X^{s_1} - 2 \varepsilon \rme^{\beta \phi} e_{\mu}{}^{s_2} m_{\nu} \rmd X^{\mu} \rmd X^{\nu} +  e_{\mu}{}^{\tilde{A}} e_{\nu}{}^{\tilde{B}} h_{\tilde{A} \tilde{B}}(\varepsilon) \rmd X^\mu \rmd X^\nu \,.
\end{align}
Upon redefining $\rme^{\beta \phi} e_{\mu}{}^{s_2} \rightarrow e_{\mu}{}^{s_2}$, this gives the following Ansatz:
\begin{align}
    \label{eq:unifiednullAnsatz}
    \rmd \hat{s}^2 &= - 2 \varepsilon e_{\mu}{}^{s_2} \rmd X^\mu \rmd X^{s_1} - 2 \varepsilon e_{\mu}{}^{s_2} m_{\nu} \rmd X^{\mu} \rmd X^{\nu} +  e_{\mu}{}^{\tilde{A}} e_{\nu}{}^{\tilde{B}} h_{\tilde{A} \tilde{B}}(\varepsilon) \rmd X^\mu \rmd X^\nu \,.
\end{align}
Since $\hat{G}_{s_1 s_1} = 0$ in this metric, we see that $X^{s_1}$ corresponds to a null direction after the $\omega \rightarrow \infty$ limit. The $(D+1)$-dimensional metric Ansatz \eqref{eq:unifiednullAnsatz} is then the one that is appropriate for 
a spacelike null reduction from signature $(1,D)$ to signature $(1,D-1)$, if $\varepsilon = -1$, and for a timelike null reduction from signature $(2,D-1)$ to signature $(1,D-1)$, in case $\varepsilon = +1$.

Let us next consider the following action for a massless particle in a $(D+1)$-dimensional spacetime with metric of the form \eqref{eq:unifiednullAnsatz}:
\begin{align}\label{mpa}
 S &= -\frac{1}{2}\int \rmd\lambda \, e^{-1}\varepsilon {\dot X}^M{\dot X}^N \hat{G}_{MN} \nonumber \\
 &= \int \rmd \lambda \, e^{-1} \left(e_{\mu}{}^{s_2} \dot{X}^{\mu} \dot{X}^{s_1} + e_{\mu}{}^{s_2} m_{\nu} \dot{X}^{\mu} \dot{X}^{\nu} - \frac{\varepsilon}{2} h_{\tilde{A}\tilde{B}}(\varepsilon) e_{\mu}{}^{\tilde{A}} e_{\nu}{}^{\tilde{B}} \dot{X}^{\mu} \dot{X}^{\nu} \right) \,,
 \end{align}
where $e$ is a worldline Einbein field. Since $e_{\mu}{}^{s_2}$, $e_{\mu}{}^{\tilde{A}}$ and $m_{\mu}$ only depend on the $X^\mu$ coordinates, the compact coordinate $X^{s_1}$ appears as a cyclic variable. Its associated momentum is thus conserved:
\begin{align}
    e^{-1} e_{\mu}{}^{s_2} \dot{X}^{\mu} =  -\varepsilon m c\,, \qquad \qquad \text{with $m$ constant} \,.
\end{align}
We can use this constraint to eliminate the worldline Einbein as
\begin{align}
    e^{-1} = -\varepsilon \frac{m c}{e_{\mu}{}^{s_2} \dot{X}^\mu} \,.
\end{align}
Substituting\,\footnote{One might object that this substitution is not consistent, since the solution for $e$ is obtained from the equation of motion of $X^{s_1}$ and not from its own equation of motion. In this case, however, it leads to the same results as those obtained by Routhian reduction. Alternatively, one can also insert \eqref{eq:rescstandardKK} in the first equation of \eqref{mpa}, use Routhian reduction to eliminate $\dot{X}^{s_1}$ and take the $\omega \rightarrow \infty$ limit. This also leads to \eqref{eq:nullredres}.} in \eqref{mpa}, one obtains:
\begin{align} \label{eq:nullredres}
    S &= m c \int \rmd \lambda \, \left( \frac{1}{2} \frac{h_{\tilde{A}\tilde{B}}(\varepsilon) e_{\mu}{}^{\tilde{A}} e_{\nu}{}^{\tilde{B}} \dot{X}^{\mu} \dot{X}^{\nu}}{e_{\mu}{}^{s_2} \dot{X}^\mu} - \varepsilon m_{\mu} \dot{X}^{\mu} - \varepsilon \dot{X}^{s_1} \right) \nonumber \\
    &= m c \int \rmd \lambda \, \left( \frac{1}{2} \frac{h_{\tilde{A}\tilde{B}}(\varepsilon) e_{\mu}{}^{\tilde{A}} e_{\nu}{}^{\tilde{B}} \dot{X}^{\mu} \dot{X}^{\nu}}{e_{\mu}{}^{s_2} \dot{X}^\mu} - \varepsilon m_{\mu} \dot{X}^{\mu}  \right) \,,
\end{align}
where in the second step we have dropped a total derivative. The dependence on the compact coordinate $X^{s_1}$ is then completely gone and the resulting action can be interpreted as that of a particle moving in $D$ spacetime dimensions. Renaming $s_2$ to $s$ and using the notation \eqref{eq:XsXAt}, we recognize it as the action \eqref{eq:BargmannAliceunified} that unifies the actions of the Bargmann and Alice particles. By choosing $\varepsilon = -1$ and $\varepsilon = +1$, we thus conclude that the latter indeed arise from the spacelike and timelike null reduction of a massless particle in a spacetime with one and two time directions, respectively. 

\section{Conclusions and Outlook} \label{sec:outlook}

In this paper, we have argued that generalized Galilean and Carrollian limits are closely related and can be treated in a unified manner. In particular, both limits can be performed by rescaling $p+1$ directions, that are interpreted as directions longitudinal to the worldvolume of a $p$-brane, with a dimensionless parameter $\omega$ and taking $\omega \rightarrow \infty$. When time is longitudinal and thus rescaled with $\omega$, this procedure sends the speed of light to infinity in the $D-p-1$ Euclidean directions transverse to the $p$-brane and corresponds to the $p$-brane Galilei limit that is known from the literature \cite{Garcia:2002fa,Brugues:2004an,Gomis:2004pw,Bergshoeff:2020xhv,Bergshoeff:2023rkk}. On the other hand, when time is transversal, the limit has the effect of sending the speed of light to zero in the $p+1$ rescaled longitudinal Euclidean directions. It is thus a generalized Carrollian limit, that we have called ``the Euclidean $p$-brane Carroll limit", in line with viewing the rescaled directions as those spanning the worldvolume of a Euclidean $p$-brane. Our treatment of generalized Carrollian limits as Euclidean brane limits differs from the original literature \cite{Cardona:2016ytk,Roychowdhury:2019aoi,Kluson:2017fam,Kluson:2022jxh,Bergshoeff:2020xhv,Bergshoeff:2023rkk}, but is suggested as appropriate in the context of string theory by recent developments on U-duality of non-Lorentzian limits of string theory \cite{Blair:2023noj,Gomis:2023eav,Blair:2024aqz,Blair:2025prd,Blair:2025nno,Argandona:2025jhg}. In ordinary string theory and supergravity, D$p$-branes occur as supersymmetric objects and the $p$-brane Galilei limits associated to their Lorentzian worldvolumes constitute interesting decoupling limits. Under timelike T-duality, these limits are mapped to generalized Carrollian limits in Hull's looking-glass land of exotic string and supergravity theories \cite{Hull:1998vg,Hull:1998ym,Hull:1998fh}. Since these theories have Euclidean supersymmetric branes, Euclidean $p$-brane Carrollian limits are the natural decoupling limits in this looking-glass land. 

The unified way of taking non-Lorentzian limits, presented in this paper, allows one to translate statements about generalized Galilean limits to Carrollian analogues, by changing the role played by time in the limit procedure. We have illustrated this for the case of $0$-brane, i.e., particle limits. We have in particular shown that there is a Carrollian counterpart of the critical nonrelativistic limit that is used to derive the Bargmann action of a nonrelativistic massive particle. This critical version of the Euclidean $0$-brane Carroll limit starts from the action of a relativistic massive tachyon and leads to a new kind of Carrollian particle, that we have called the Alice particle. The Alice particle is a non-trivial realization of a centrally extended symmetry algebra, that is a Carrollian variant of the Bargmann algebra, i.e., the central extension of the Galilei algebra. It is known that the Bargmann particle in $D$ spacetime dimensions can be obtained from null reduction of the massless particle action in a relativistic spacetime with signature $(1,D)$ \cite{Duval:1984cj}. This null reduction is spacelike in the sense that it can be viewed as a limit of reductions over a spacelike circle \cite{Seiberg:1997ad,Sen:1997we}. In this paper, we have argued that there also exists a null reduction that is timelike, i.e., a limit of reductions over a timelike circle and that allows one to obtain the Alice particle in $D$ spacetime dimensions from the relativistic massless particle in signature $(2,D-1)$.

The 10-dimensional Bargmann particle action plays a special role in type IIA string theory and its strongly coupled version, M-theory, where it is the nonrelativistic limit of the action of a single D$0$-brane. Generalizing to $N$ super-D$0$-branes (with $N > 1$), this limit is recognized \cite{Blair:2024aqz,Blair:2025prd,Guijosa:2025mwh} as the decoupling limit that leads to the BFSS matrix model \cite{Banks:1996vh,Susskind:1997cw}. When compactifying 11-dimensional M-theory on a spatial circle to type IIA string theory, D$0$-branes correspond to Kaluza-Klein excitations of the 11-dimensional supergravity multiplet that have positive momentum along the compact direction. In the limit that turns spatial circle reductions into the spacelike null reduction and that leads to the Bargmann particle or BFSS matrix model, all states with negative or zero Kaluza-Klein momentum decouple, and only D$0$-brane states survive in the spectrum. This is key to conjecturing that the large $N$-limit of the BFSS matrix model gives a non-perturbative formulation of M-theory. Our results for the Alice particle show that, in 10 dimensions, it can be viewed as a Carrollian version of the Bargmann particle in IIA${}^*$ string theory (that lives in signature $(1,9)$) and its strong coupling limit, namely M${}^*$ theory in signature $(2,9)$. In particular, 
our derivation of the Alice particle as a Euclidean $0$-brane limit of a relativistic massive tachyon agrees with its interpretation as a decoupling limit of a D$0{}^*$-brane, which is Euclidean (i.e., tachyonic). Likewise, our timelike null reduction procedure that gives the 10-dimensional Alice particle, corresponds to a limit of M${}^*$ theory on timelike circles, in which all states, apart from the D$0{}^*$-branes that have positive Kaluza-Klein momentum, decouple from the spectrum. A generalization of the action of a single Alice particle to $N$ coinciding Alice superparticles, is then expected to correspond to a BFSS${}^*$ matrix model, whose large $N$-limit gives a non-perturbative description of M${}^*$ theory \cite{Blair:2025nno,Argandona:2025jhg}. The Bargmann and Alice particles are thus natural objects in type IIA and IIA${}^*$ string theory respectively. This is strengthened by the observation we made in Section \ref{sec:BargmannAlice} that the Bargmann particle is stable in the presence of a negative cosmological constant, while the Alice particle requires a positive cosmological constant for stability.

Our work leads to various research directions that we wish to explore in the future. Here, we restricted ourselves to $0$-brane limits of particle actions. In a follow-up paper \cite{Alicestrings}, we will extend our work on the Alice particle to critical Carrollian limits of worldvolume actions of Euclidean strings and branes in the looking-glass land of exotic string theories obtained via timelike T-duality. The resulting Carrollian string action should give rise to a Carrollian analogue of nonrelativistic string theory \cite{Gomis:2000bd,Danielsson:2000gi,Danielsson:2000mu}, while supersymmetric and nonabelian versions of Carrollian D$p$-brane actions are expected to give the field theory sides of examples of the dS/CFT correspondence \cite{Argandona:2025jhg,Blair:2025nno}. It would also be interesting to include supersymmetry in our analysis, so that we can discuss Euclidean Carroll limits of superstrings, super-D$p{}^*$-branes and supergravity theories in the looking-glass land of exotic string theories. In this regard, one should note that string and supergravity theories that are obtained via timelike T-duality are not invariant under usual supersymmetries, but rather under a twisted form thereof that is characterized by noncompact R-symmetry groups. It has recently been shown in 3 dimensions that theories with twisted supersymmetry admit so-called magnetic Carroll limits, in which the anti-commutators of the supercharges give both time and spatial translations \cite{Bulunur:2026yav}. Another potential research topic stems from the observation that one can take the critical limit of the action of a particle moving in a Schwarzschild black hole metric, similar to how we included a cosmological constant in the Bargmann and Alice particle actions in Section \ref{sec:BargmannAlice}. In the exterior region of the black hole, this limit leads to a Bargmann particle, moving in a flat background, but coupled to the Newton potential of a massive point mass. In the interior of the black hole, the roles of time and space are interchanged and consequently, the limit results in an Alice particle, moving in a non-trivial, time-dependent background and coupled to a non-trivial potential. Perhaps the Alice particle can thus play a part in investigations of black hole interiors. Finally, it is well-known that one can obtain 4-dimensional Newton-Cartan gravity, a diffeomorphism-covariant version of Newtonian gravity, from spacelike null reduction of 5-dimensional General Relativity. The timelike null reduction that we discussed in this paper, can be used to derive a Carrollian version of Newton-Cartan gravity, starting from General Relativity in a 5-dimensional spacetime with signature $(2,3)$. It would be interesting to perform this null reduction and study the properties of the resulting Carrollian gravity theory. 

Admittedly, the Alice particle, as a limit of a tachyon or as a null reduction of a massless particle in a spacetime with two times, has rather unconventional features. Nevertheless, it may perhaps serve as a useful guide to explore the various squares on the chessboard of string theory's exotic looking-glass land and help further clarify their role in quantum gravity and holography. We hope to report on the Alice particle and the research lines outlined in the above paragraph in the future, after the passage of some time(s).

\section*{Acknowledgements}

We would like to thank Jos\'e Figueroa-O'Farrill, Chris Hull, Neil Lambert, Prahar Mitra, Rishi Mouland, Niels Obers, Marios Petropoulos, Massimo Porrati, Stefan Prohazka, Joseph Smith, Andrew Strominger and Ziqi Yan for useful discussions. 
The work of L.R. has been supported by the
Ramon y Cajal fellowship RYC2023-042671-I, funded by MCIU/AEI/10.13039/501100011033 and
FSE+, by the Spanish Ministry of Science and Innovation (MCI), the State Research Agency (AEI),
and the European Regional Development Fund (FEDER, EU) under the grant PID2024-155685NB-
C22 and by Fundacion Seneca de la Region de Murcia FSRM/10.13039/100007801 (22581/PI/24). 
The work of E.S.F is supported by the fellowship FPI-UM R-803/2024 of the University of Murcia, by the Spanish Ministry of Science and Innovation (MCI), the State Research Agency (AEI), and the European Regional Development Fund (FEDER, EU) under the grant PID2024-155685NB-C22 and by Fundación Séneca de la Región de Murcia FSRM/10.13039/100007801 (22581/PI/24). The work of S.Z.~is supported by an HrZZ MOBDOK-2023 fellowship of the Croatian Science Foundation. The work of E.B., J.R.~and S.Z.~is supported by the Croatian Science Foundation project IP-2022-10-5980 “Non-relativistic supergravity and applications” and by the European Union–NextGenerationEU. The views and opinions expressed are those of the authors only and do not necessarily reflect the official views of the European Union or the European Commission. Neither the European Union nor the European Commission can be held responsible for them. The work of S.Z.~was supported by a short term scientific mission grant from the COST action CA22113 THEORY-CHALLENGES.

\appendix 

\section{Schr\"odinger and Euclidean Carroll-Schr\"odinger Lagrangians} \label{sec:SchrCarrSchr}

In this Appendix, we will show that the unified critical limit that leads to the Bargmann and Alice particles can also be applied to the Lagrangian of a massive complex scalar that is minimally coupled to a one-form potential. In the Galilean case, this limit leads to the Schr\"odinger Lagrangian, whereas in the Carrollian case, one obtains a higher-dimensional generalization of the Carroll-Schr\"odinger Lagrangian that in two spacetime dimensions was studied in \cite{Najafizadeh:2024imn,Rojas:2025rot,Rojas:2025ygg}. We refer to this generalization as ``the Euclidean Carroll-Schr\"odinger Lagrangian".

Our starting point is the following Lagrangian for a complex scalar field $\Phi$, minimally coupled to a Lorentzian background metric $G_{\mu\nu}$ and one-form gauge field $M_\mu$:
\begin{gather}
    \label{eq:LKGt}
    E^{-1} \mathcal{L} = -\frac12 G^{\mu\nu} D_{\mu} \Phi D_{\nu} \Phi^* +  \frac{1}{2}\varepsilon \big(Mc\big)^2 \Phi \Phi^* \,,
\end{gather}
where the covariant derivative $D_\mu\Phi$ is defined by
\begin{align}
        D_{\mu} \Phi \equiv \partial_{\mu} \Phi - \rmi Mc M_{\mu} \Phi \,.
\end{align}
Note that for $\varepsilon = 1$, the field $\Phi$ is tachyonic, whereas for $\varepsilon = -1$, the Lagrangian \eqref{eq:LKGt} is tachyon-free. Writing the metric $G_{\mu\nu}$ in terms of a Vielbein $E_{\mu}{}^{A}$ with inverse $E^{\mu}{}_{A}$ and decomposing the flat index $A$ into $s$ and $\tilde{A}$ as in \eqref{redef11}, we can insert the redefinitions 
\begin{alignat}{4}
E_\mu{}^s &= \omega \, e_\mu{}^s\,, \qquad  & E_\mu{}^{\tilde A} &= e_\mu{}^{\tilde A}\,, \qquad & E^{\mu}{}_s &= \omega^{-1} \, e^{\mu}{}_s \,, \qquad & E^{\mu}{}_{\tilde{A}} &= e^{\mu}{}_{\tilde{A}} \,, \nonumber \\
 M_\mu &= \omega\, e_\mu{}^s + \omega^{-1}\,m_\mu\,, \qquad & & & & & M &= \omega\, m\,,
\end{alignat}
in \eqref{eq:LKGt} and take the $\omega\to \infty$ limit. The terms that are of leading quadratic order in $\omega$ cancel and we obtain the following Lagrangian
\begin{align}
    \label{eq:LKGtlim}
    e^{-1} \mathcal{L}=-\frac{\rmi}{2}\varepsilon\, m c\,  e^{\mu}{}_s \big(\Phi^*\, \tilde D_\mu \Phi - \Phi\, \tilde D_\mu \Phi^*\big) - \frac12 e^{\mu \tilde{A}} e^{\nu \tilde{B}} h_{\tilde{A} \tilde{B}}(\varepsilon) \tilde D_{\mu} \Phi \,  \tilde D_{\nu} \Phi^* \,,
\end{align}
where $e^{-1}\equiv \det$ ($e^\mu{}_s\,, e^\mu{}_{\tilde A})$ and the covariant derivative is given by
\begin{equation}
 \tilde D_\mu \Phi=\partial_\mu \Phi - \rmi \, m c \, m_\mu \Phi\,.
 \end{equation}
Let us consider the case $\varepsilon=-1$ (Galilei) first. Choosing $m_{\mu} = 0$ and flat space with inertial coordinates $x^{\mu} = (x^s = c t,\, x^{\tilde{A}})$,\,\footnote{In these coordinates, the flat space inverse Vielbein $e^{\mu}{}_s$ is given by $e^{t}{}_{s} = 1$ and $e^{\tilde{A}}{}_{s} = 0$. For $e^{\mu}{}_{\tilde{A}}$, one has $e^{t}{}_{\tilde{A}} = 0$ and $e^{\tilde{B}}{}_{\tilde{A}} = \delta^{\tilde{B}}_{\tilde{A}}$.} the Lagrangian \eqref{eq:LKGtlim} becomes the standard Schr\"odinger Lagrangian:
\begin{align}
    \label{eq:Schr}
    \mathcal{L}_{\text{Schr\"odinger}} &= \frac{\rmi}{2} m \left(\Phi^* \partial_t \Phi - \Phi \partial_t \Phi^*\right) - \frac12 \delta^{\tilde{A}\tilde{B}} \partial_{\tilde{A}} \Phi \partial_{\tilde{B}} \Phi^* \,.
\end{align}
For $\varepsilon = +1$ (Carroll), picking $m_\mu = 0$ and flat space with inertial coordinates $x^{\mu} = (x^{\tilde{A}}, x^s = y)$,\,\footnote{In these coordinates, the flat space inverse Vielbein $e^{\mu}{}_s$ is given by $e^{\tilde{A}}{}_{s} = 0$ and $e^{y}{}_{s} = 1$. For $e^{\mu}{}_{\tilde{A}}$, one has $e^{\tilde{B}}{}_{\tilde{A}} = \delta^{\tilde{B}}_{\tilde{A}}$ and $e^{y}{}_{\tilde{A}} = 0$.} one finds that \eqref{eq:LKGtlim} is given by:
\begin{align}
    \label{eq:CarrSchr}
    \mathcal{L}_{\text{Eucl. Carroll-Schr\"odinger}} &= - \frac{\rmi}{2} p \left(\Phi^* \partial_y \Phi - \Phi \partial_y \Phi^* \right) - \frac12 \eta^{\tilde{A}\tilde{B}} \partial_{\tilde{A}} \Phi \partial_{\tilde{B}} \Phi^* \,,
\end{align}
where we have put $m c = p$. In two spacetime dimensions, this theory was studied in \cite{Najafizadeh:2024imn,Rojas:2025rot,Rojas:2025ygg}, where it was called the ``Carroll-Schr\"odinger Lagrangian". The Lagrangian \eqref{eq:CarrSchr} is a generalization of this Carroll-Schr\"odinger theory that is defined in any spacetime dimension.
The Schr\"odinger \eqref{eq:Schr} and Euclidean Carroll-Schr\"odinger \eqref{eq:CarrSchr} Lagrangians are invariant under the Bargmann and Alice algebras, respectively. In particular, the Bargmann transformations that leave the Schr\"odinger Lagrangian \eqref{eq:Schr} invariant, act on $\Phi$ as follows:
\begin{align}
    \label{eq:BargtrafoPhi}
    \delta \Phi &= \zeta_{\text{B}}\, \partial_t \Phi + \zeta_{\text{B}}{}^{\tilde{A}} \, \partial_{\tilde{A}} \Phi - \left(\lambda_{\text{B}}\right){}^{\tilde{A}}{}_{\tilde{B}}\, x^{\tilde{B}} \,\partial_{\tilde{A}} \Phi - \lambda_{\text{B}}{}^{\tilde{A}}\, t \, \partial_{\tilde{A}} \Phi + \rmi\, m \, \lambda_{\text{B}}{}^{\tilde{A}}\, x^{\tilde{B}} \, \delta_{\tilde{A}\tilde{B}} \, \Phi - \rmi\, m\, \sigma_{\text{B}} \, \Phi \,.
\end{align}
Here, $\zeta_{\text{B}}$, $\zeta_{\text{B}}{}^{\tilde{A}}$, $\left(\lambda_{\text{B}}\right){}^{\tilde{A}}{}_{\tilde{B}}$, $\lambda_{\text{B}}{}^{\tilde{A}}$ and $\sigma_{\text{B}}$ are the (constant) parameters of time translations, spatial translations, spatial rotations, Galilean boosts and the central charge transformation of the Bargmann algebra respectively. Similarly, the action on $\Phi$ of the Alice transformations that leave the Euclidean Carroll-Schr\"odinger Lagrangian \eqref{eq:CarrSchr} invariant, is given by:
\begin{align}
    \label{eq:AlicetrafoPhi}
    \delta \Phi &= \zeta_{\text{A}} \, \partial_y \Phi + \zeta_{\text{A}}{}^{\tilde{B}} \, \partial_{\tilde{B}} \Phi - \left(\lambda_{\text{A}}\right){}^{\tilde{B}}{}_{\tilde{C}}\, x^{\tilde{C}} \, \partial_{\tilde{B}} \Phi - \lambda_{\text{A}}{}^{\tilde{B}} \, y \, \partial_{\tilde{B}} \Phi - \rmi\, p \, \lambda_{\text{A}}{}^{\tilde{B}} \, x^{\tilde{C}} \, \eta_{\tilde{B}\tilde{C}} \, \Phi + \rmi \, p \, \sigma_{\text{A}} \, \Phi \,,
\end{align}
where $\zeta_{\text{A}}$, $\zeta_{\text{A}}{}^{\tilde{B}}$, $\left(\lambda_{\text{A}}\right){}^{\tilde{B}}{}_{\tilde{C}}$, $\lambda_{\text{A}}{}^{\tilde{B}}$ and $\sigma_{\text{A}}$ are the (constant) parameters of the $y$-translation, longitudinal translations, longitudinal Lorentz transformations, Carrollian boosts and central charge of the Alice algebra. Using the transformation rules \eqref{eq:BargtrafoPhi} and \eqref{eq:AlicetrafoPhi}, one can explicitly check that the commutators \eqref{Bargmann} that give the central charge of the Bargmann and Alice algebras, are non-trivially realized on $\Phi$. Let us note that the Schr\"odinger equation \eqref{eq:Schr} is invariant under a larger set of symmetry transformations than the Bargmann ones given in \eqref{eq:BargtrafoPhi}. In particular, it is invariant under anisotropic dilatations and special conformal transformations that extend the Bargmann to the Schr\"odinger algebra. It is natural to expect that a similar conformal extension exists for the Alice algebra. In this case, the anisotropic dilatations should rescale the $y$-coordinate with weight $2$ and the remaining coordinates with weight $1$, as is indicated by the fact that there is only one derivative with respect to the former, and two derivatives with respect to the latter in the action \eqref{eq:CarrSchr}.

Let us contrast the above critical Bargmann and Alice limits of a massive complex scalar to various noncritical Galilei and (ordinary) Carroll limits of real scalar field theories that have appeared in the literature \cite{Souriau1997,Duval:2014uoa,Bergshoeff:2014jla,Bergshoeff:2015wma,deBoer:2021jej,Bergshoeff:2022eog,Henneaux:2021yzg,Ecker:2024czx,Figueroa-OFarrill:2023vbj,Figueroa-OFarrill:2023qty,Figueroa-OFarrill:2024ocf}. Dropping our unified notation and following \cite{Bergshoeff:2022eog}, we  consider the following Lagrangian for a real scalar field:
\begin{align}\label{case1}
 E^{-1} \,\mathcal{L}_{\rm rel} = +\frac12\, E^{\mu}{}_{0} E^{\nu}{}_{0} \partial_\mu \Phi \partial_\nu \Phi  - \frac12\, E^{\mu}{}_{\tilde A} E^{\nu}{}_{\tilde B} \delta^{\tilde A\tilde B} \partial_\mu \Phi \partial_\nu \Phi -\frac12 \alpha M^2 \Phi^2\,,
\end{align}
where $\tilde A = 1, \cdots , D-1$ and $\alpha$ is a sign parameter that indicates whether the field $\Phi$ is tachyonic ($\alpha = -1$) or non-tachyonic ($\alpha = +1$). Substituting the Galilei redefinitions $E^{\mu}{}_{0} = \omega^{-1} \tau^{\mu}\,, E^{\mu}{}_{\tilde A} = e^{\mu}{}_{\tilde{A}}$,  we obtain
\begin{align}
 e^{-1} \,\mathcal{L}_{\rm rel} = +\frac{1}{2\omega^2} \tau^\mu\tau^\nu \partial_\mu \Phi \partial_\nu \Phi  -\frac12\, e^{\mu}{}_{\tilde{A}} e^{\nu}{}_{\tilde{B}} \delta^{\tilde{A}\tilde{B}} \partial_\mu \Phi \partial_\nu \Phi -\frac12 \alpha M^2 \Phi^2\,,
\end{align}
where $e^{-1} = \det (\tau^\mu,e^{\mu}{}_{\tilde{A}})$ and we have ignored an overall power of $\omega$, since it can be absorbed in a further redefinition of $\Phi$.
There are now two ways to proceed. First, by  choosing  $\alpha=-1$ and taking the limit $\omega \to \infty$, we obtain the following `magnetic' Galilei Lagrangian:
\begin{align}
 e^{-1} \,\mathcal{L}_{\rm magnetic\ Galilei} =  - \frac12\, e^{\mu}{}_{\tilde A} e^{\nu}{}_{\tilde B} \delta^{\tilde A\tilde B} \partial_\mu \Phi \partial_\nu \Phi +\frac12 M^2 \Phi^2\,.
\end{align}
Second, we first redefine $\Phi = \omega\, \phi, M = \omega^{-1}\, m$ and control the resulting $\omega^2$--divergence by applying a Hubbard-Stratonovich transformation introducing auxiliary scalars $\chi^{\tilde A}$. Next, choosing $\alpha = +1$ and taking  the limit $\omega\to \infty$, the auxiliary fields $\chi^{\tilde A}$ become Lagrange multipliers and we obtain the following Lagrangian:\,\footnote{The effect of the Hubbard-Stratonovich transformation is to set the leading divergence to zero by a Lagrange multiplier and to remain with the sub-leading terms.}
\begin{align}
 e^{-1} \,\mathcal{L}_{\rm electric\ Galilei} = +\frac12\, \tau^\mu\tau^\nu \partial_\mu \phi \partial_\nu \phi  + \chi^{\tilde{A}}e^\mu{}_{\tilde{A}}\partial_\mu\phi -\frac12 m^2 \phi^2\,.
 \end{align}

In the Carroll case, we start from the same relativistic Lagrangian \eqref{case1} but now perform the standard Carroll redefinitions $E^{\mu}{}_{0} = \omega \, \tau^\mu\,, E^{\mu}{}_{\tilde A} = e^{\mu}{}_{\tilde A}$. In this way we obtain
\begin{align}
 e^{-1} \,\mathcal{L}_{\rm rel} = +\frac12\, \omega^2 \tau^\mu\tau^\nu \partial_\mu \Phi \partial_\nu \Phi  -\frac12\, e^{\mu}{}_{\tilde A} e^{\nu}{}_{\tilde B}\delta^{\tilde A\tilde B} \partial_\mu \Phi \partial_\nu \Phi -\frac12\alpha  M^2 \Phi^2\,.
\end{align}
Again, there are two options. First, to deal with the quadratic
divergence in the first term, we apply a Hubbard-Stratonovich transformation introducing an auxiliary field $\chi$. 
Next, choosing $\alpha=-1$ and taking the limit $\omega \to \infty$,  $\chi$  becomes a Lagrange multiplier, and we obtain the following magnetic Carroll Lagrangian:
\begin{align}\label{case3b}
 e^{-1} \,\mathcal{L}_{\rm magnetic \ Carroll} = - \chi \tau^\mu\partial_\mu\Phi  -  \frac12\, e^{\mu}{}_{\tilde A} e^{\nu}{}_{\tilde B}\delta^{\tilde A\tilde B} \partial_\mu \Phi \partial_\nu \Phi +\frac12 M^2 \Phi^2\,.
\end{align}
Second, we first redefine $\Phi =\omega^{-1}\,\phi, M=\omega\, m$. We  then choose $\alpha=+1$ and take  the limit $\omega \to \infty$ after which we obtain the following Lagrangian:
\begin{align}\label{eq:case3c}
 e^{-1} \,\mathcal{L}_{\rm electric\ Carroll } = +\frac12\,  \tau^\mu\tau^\nu \partial_\mu \phi \partial_\nu \phi -\frac12 m^2 \phi^2\,.
\end{align}
This finishes our overview of the noncritical limits. Note that in all these cases we end up with a Lagrangian whose symmetry algebra is given by either the Galilei or (ordinary) Carroll algebra with no central extension. 

\section{Dispersion Relations} \label{sec:disprels}

In this Appendix, we will mimic the critical $0$-brane Galilei and Euclidean $0$-brane Carroll limits simultaneously, at the level of the dispersion relation for the states of a relativistic bradyon or tachyon. This gives dispersion relations of the Bargmann and Alice particles, respectively. Moreover, it allows one to explicitly see that part of the spectrum of states of the bradyon or tachyon decouples when taking these critical limits. 

To start, we write the relativistic dispersion relation of a bradyon or tachyon with mass $M$ in the unified notation adopted in this paper, as follows: 
\begin{equation}
  \label{eq:dispersion0}
  \varepsilon \big(P^s\big)^2 + P^{\tilde A}P^{\tilde B}  h_{\tilde A\tilde B}(\varepsilon) -\varepsilon\big(M c\big)^2 = 0\,,
\end{equation}
For $\varepsilon=-1$ and $s$ the time direction ($0$-brane Galilei), this is the dispersion relation of a relativistic bradyon, while for $\varepsilon=+1$ and $s$ a spatial direction (Euclidean $0$-brane Carroll), we are dealing with a tachyon. To take a critical limit of this dispersion relation, we couple the particle to a background gauge field, whose only non-zero component is in the $s$ direction and given by the constant value $M c$. The effect of this background gauge field is to shift $P^s$ in the first term of \eqref{eq:dispersion0} with $M c$, and we thus consider the following dispersion relation:
\begin{equation}
    \label{eq:dispersion1}
    \varepsilon \big(P^s + M c \big)^2 + P^{\tilde A}P^{\tilde B}  h_{\tilde A\tilde B}(\varepsilon) -\varepsilon\big(M c\big)^2 = 0\,.
\end{equation}
We then insert the following redefinitions:
\begin{equation}
P^s = \omega^{-1}\, p^s\,,\hskip 1truecm P^{\tilde A} = p^{\tilde A}\,, \hskip 1truecm M =\omega\, m \,,  
\end{equation}
to obtain:
\begin{equation}
    \varepsilon\big(\omega^{-1} \, p^s + \omega \, m\, c\big)^2 + p^{\tilde A}p^{\tilde B}  h_{\tilde A\tilde B}(\varepsilon) - \varepsilon\, \omega^2 \, \big(m\, c\big)^2 = 0\,.
\end{equation}
This equation has 2 solutions for $p^s$:
\begin{align}
    p^s &= - \frac12 \varepsilon \frac{p^{\tilde{A}} p^{\tilde{B}} h_{\tilde{A}\tilde{B}}(\varepsilon)}{m c} + \mathcal{O}(\omega^{-2}) \,, \qquad \text{and} \nonumber \\
    p^s &= -2 \omega^2 m c + \frac12 \varepsilon \frac{p^{\tilde{A}} p^{\tilde{B}} h_{\tilde{A}\tilde{B}}(\varepsilon)}{m c} + \mathcal{O}(\omega^{-2}) \,.
\end{align}
The second solution diverges in the $\omega \rightarrow \infty$ limit and thus corresponds to states that decouple from the spectrum in the limit. The first solution has a well-defined $\omega \rightarrow \infty$ limit and the states associated to it remain in the spectrum and obey the following dispersion relation after the limit:
\begin{equation}
p^s = -\frac12 \varepsilon \frac{ p^{\tilde A}p^{\tilde B}  h_{\tilde A\tilde B}(\varepsilon)}{m c}\,.
\end{equation}
Specfiying $\varepsilon = -1$ and $\varepsilon = +1$, we find the following two dispersion relations:
\begin{align}
    \label{eq:BargAldisp}
    {\varepsilon = -1 \ \ (0\mathrm{-brane\ Galilei})\,:}\ \ \ 
& E \equiv p^s c  = \frac{1}{2 m} p^{\tilde A}p^{\tilde B}  \delta_{\tilde A\tilde B} > 0\,,\cr
{\varepsilon = +1 \ \ (\mathrm{Eucl.}\ 0\mathrm{-brane\ Carroll})\,:}\ \ \ 
& p^y \equiv p^s   = -\frac{1}{2 p} p^{\tilde A}p^{\tilde B}  \eta_{\tilde A\tilde B} \,. 
\end{align}
where we have set $m c = p$ for $\varepsilon = +1$. These can be identified as dispersion relations of the Bargmann and Alice particles, respectively. Indeed, identifying the momenta $p^{\tilde{A}}$ as in \eqref{eq:momenta}, $E$ is the standard nonrelativistic energy, i.e., the conserved Noether charge associated to translations of the worldline parameter $t$ in the Bargmann particle action \eqref{eq:Bargmann}. Similarly, $p^y$ is a conserved charge associated to translations of the worldline parameter $y$ in the Alice particle action \eqref{eq:Alice}. The above argument explicitly shows that in the $0$-brane Galilei limit anti-particles decouple from the spectrum. Analogously, in the Euclidean $0$-brane Carroll limit, it is either left- or right-moving (in the $y$-direction) particles that decouple. The dispersion relations \eqref{eq:BargAldisp} are also found by Fourier transforming the Schr\"odinger and Euclidean Carroll-Schr\"odinger equations \eqref{eq:Schr} and \eqref{eq:CarrSchr}.

Let us, for completeness, also show how the dispersion relations of the noncritical Lagrangian limits discussed in the previous Appendix can also be obtained via a similar noncritical limit of relativistic dispersion relations. Whenever we apply a Hubbard-Stratonovich transformation in the Lagrangian limit, we impose a constraint in the dispersion relation. To compare with Appendix \ref{sec:SchrCarrSchr} and the literature, we give the same names to the various limits and drop our unified notation. An early example of a noncritical magnetic  Galilean limit of a dispersion relation describing a relativistic tachyon was given in \cite{Batlle:2017cfa} (see also \cite{Aldaya:2024acv}). Substituting the redefinition $P^0  = \omega^{-1}\, p^0$ and $\vec P=\vec p$ into the tachyonic dispersion relation
\begin{equation}\label{tachyonic}
{\rm tachyon\,:}\ \ \ -\big(P^0\big)^2 + {\vec P}^2 -\big(Mc\big)^2 = 0 \,,
\end{equation}
and taking the $\omega\to \infty$ limit, one obtains
\begin{equation}
 {\rm magnetic\ Galilean\,:}\ \ \  
 {\vec p}^2 = k^2\,,
   \end{equation}
where $k\equiv Mc$ is the ``colour''/momentum of a massless Galilean particle with zero energy.  The electric Galilean limit is obtained by imposing the constraint  $\vec P=0$ and substituting  the redefinitions  $P^0  = \omega^{-1}\, p^0$ and $M = \omega^{-1}\, m$ into the dispersion relation of a bradyon
\begin{equation}\label{particle}
{\rm massive\ particle\,:}\ \ \ -\big(P^0\big)^2 + {\vec P}^2 +\big(Mc\big)^2 = 0\,.
\end{equation}
 Taking the $\omega \rightarrow \infty$ limit, one gets
 \begin{equation}
     {\rm electric\ Galilean\,:}\ \ \ (p^0)^2 = m^2\,.
 \end{equation}  
   
Similar noncritical Carroll limits have been considered in \cite{deBoer:2021jej}. These authors considered a so-called `magnetic' Carroll limit of the tachyonic dispersion relation \eqref{tachyonic}. Inserting the redefinitions $P^0  = \omega\, p^0\,, \vec P=\vec p$ and imposing the constraint $P^0=0$, one obtains after taking the $\omega\to \infty$ limit:
\begin{equation}
{\rm magnetic\ Carroll\,:}\ \ \ \vec{p}^{\, \, 2} = \big(Mc\big)^2\,.
\end{equation}
This dispersion relation corresponds to a Carroll particle with zero energy that can move. One may also consider an electric Carroll limit where, beyond the redefinitions in the previous case, one makes the additional redefinitions $M=\omega m$. Substituting these redefinitions into the particle dispersion relation \eqref{particle} and taking the $\omega\to \infty$ limit, one gets
\begin{equation}
{\rm electric\ Carroll\,:}\ \ \ (p^0)^2 = \big(mc\big)^2\,.
\end{equation}
This corresponds to a Carroll particle with non-zero energy that, however, has $\vec P = 0$ and thus can not move. All the above examples are based upon an underlying  Galilei or Carroll algebra with zero central extension. This should be contrasted with the critical limits discussed at the beginning of this Appendix, that lead to the Bargmann and Alice algebras that have a central extension.


\providecommand{\href}[2]{#2}\begingroup\raggedright\endgroup

\end{document}